\documentclass{aa}  
\usepackage{graphicx}
\usepackage{txfonts}
\usepackage{booktabs}
\begin{document}

   \title{ALMA and VLA reveal the lukewarm chromospheres of the nearby red supergiants Antares and Betelgeuse}

   \author{E. O'Gorman
          \inst{1}\fnmsep\thanks{ogorman@cp.dias.ie},
         G. M. Harper\inst{2},
          K. Ohnaka\inst{3},
          A. Feeney-Johansson\inst{1},
          K. Wilkeneit-Braun\inst{4},
          A. Brown\inst{2},
          E. F. Guinan\inst{5},
          J. Lim\inst{6},
          A. M. S. Richards\inst{7},
          N. Ryde\inst{8}
          \and
          W. H. T. Vlemmings\inst{9}
          }

   \institute{Dublin Institute for Advanced Studies, 31 Fitzwilliam Place, Dublin 2, Ireland
         \and
             Center for Astrophysics and Space Astronomy, University of Colorado, 389 UCB, Boulder, CO 80309, USA
            \and
                 Instituto de Astronom\'{ı}a, Universidad Ca\'{o}lica del Norte, Avenida Angamos 0610, Antofagasta, Chile
                 \and
            Hamburger Sternwarte, Universit\"{a}t Hamburg, Gojenbergsweg 112, 21029 Hamburg, Germany
         \and
         Department of Astrophysics and Planetary Science, Villanova University, Villanova, PA 19085, USA
         \and
         Department of Physics, The University of Hong Kong, Pokfulam Road, Hong Kong
         \and
         Jodrell Bank Centre for Astrophysics, Department of Physics and Astronomy, University of Manchester, Manchester M13 9PL, UK
         \and
         Lund University, Lund, Sweden
         \and
         Department of Space, Earth and Environment, Chalmers University of Technology, Onsala Space Observatory, 439 92, Onsala, Sweden
             }

   \date{Received February 17, 2020; Accepted May 5, 2020}

  \abstract 
   {We first present spatially resolved ALMA and VLA continuum observations of the early-M red supergiant Antares to search for the presence of a chromosphere at radio wavelengths. We resolve the free-free emission of the Antares atmosphere at 11 unique wavelengths between 0.7\,mm (ALMA band 8) and 10\,cm (VLA S band). The projected angular diameter is found to continually increase with increasing wavelength, from a low of 50.7\,mas at 0.7\,mm up to a diameter of 431\,mas at 10\,cm, which corresponds to 1.35 and 11.6 times the photospheric angular diameter, respectively. All four ALMA measurements show that the shape of the atmosphere is elongated, with a flattening of 15\% at a similar position angle. The disk-averaged gas temperature of the atmosphere initially rises from a value of 2700\,K at 1.35\,$R_{\star}$ (i.e., 0.35\,$R_{\star}$ above the photosphere) to a peak value of 3800\,K at $\sim$2.5\,$R_{\star}$, after which it then more gradually decreases to 1650\,K at 11.6\,$R_{\star}$. The rise in gas temperature between 1.35\,$R_{\star}$ and $\sim$2.5\,$R_{\star}$ is evidence for a chromospheric temperature rise above the photosphere of a red supergiant. We detect a clear change in the spectral index across the sampled wavelength range, with the flux density ${S_{\nu}} \propto {\nu}^{1.42}$ between 0.7\,mm and 1.4\,cm, which we associate with chromosphere-dominated emission, while the flux density ${S_{\nu}} \propto {\nu}^{0.8}$ between 4.3\,cm and 10\,cm, which we associate with wind-dominated emission. We show that the Antares MOLsphere is transparent at our observed wavelengths, and the \textit{\textup{lukewarm}} chromosphere that we detect is therefore real and not just an average of the cool MOLsphere and hot ultraviolet emitting gas. We then perform nonlocal thermal equilibrium modeling of the far-ultraviolet radiation field of another early-M red supergiant, Betelgeuse, and find that an additional hot (i.e., $>7000\,$K) chromospheric  photoionization component with a much smaller filling factor must also exist throughout the chromospheres of these stars.}

  \keywords{Stars: atmospheres --
                Stars: chromospheres --
                Stars: imaging --
                Stars: massive --
                Radio continuum: stars --
                Submillimeter: stars
               }
               
   \titlerunning{Lukewarm chromospheres of red supergiants}
   \authorrunning{E. O'Gorman et al.}
   \maketitle
%

\section{Introduction}\label{sec1}
All late spectral type (cool) stars are believed to contain regions in their atmospheres exterior to the photosphere where plasma is heated to temperatures above the prediction of radiative equilibrium \citep{schrijver_2000}. This region is known as the chromosphere and is formed by the dissipation of acoustic and magnetic waves, which are a consequence of the turbulent convective motions beneath the photosphere. The resulting nonradiative heating leads to a chromospheric signature that is most conspicuous at ultraviolet (UV) and radio (i.e., centimeter to submillimeter) wavelengths \citep[][and references therein]{linsky_2017}. 

The presence and nature of chromospheres in K and early-M supergiants has important implications for both deriving their highly uncertain mass-loss rates, which themselves are key inputs for stellar evolution codes \citep{meynet_2015}, and understanding the details of the as yet undeciphered mass-loss mechanisms. Moreover, because K and early-M supergiants form the majority of the red supergiants (RSGs) \citep{levesque_2017}, understanding the nature of chromospheres in these types of stars becomes pressing. Mass-loss rates derived from modeling molecules, especially CO low-J pure-rotation lines, and silicate dust models are of limited reliability (e.g., see \citealt{josselin_2000}). The RSG chromospheric far-ultraviolet (FUV) radiation field photodissociates molecules ejected from the photosphere, which in turn inhibits dust formation. This makes the total mass-loss rates unreliable because the CO/H fraction and the gas-to-dust mass ratio are poorly constrained. 

Optical-UV ($\lambda\sim 3200\,\AA$) Fe\,II emission lines have long revealed the presence of chromospheric emission from cool evolved stars \citep{herzberg_1948, boesgaard_1976}, including the RSGs. However, with the exception of Betelgeuse, little is known about chromospheric {\it \textup{struc}\textup{ture}} or the intrinsic FUV radiation field that is important for circumstellar chemistry in RSGs. Early empirical radio studies (see \citealt{harper_2001}, and references therein) and theoretical Alfv\'en wave-driven wind models led to a picture of a hot (7,000-9,000\,K) chromosphere extending out to $\sim 6$\,R$_\star$  \citep{hartmann_1984, skinner_1997}. This was in good agreement with spatially resolved UV observations made with the Hubble Space Telescope (HST) \citep{gilliland_1996, uitenbroek_1998} and radio observations lacking spatial information  \citep{newell_1982, hjellming_1983}. However, spatially resolved multiwavelength radio observations of Betelgeuse by \cite{lim_1998} found no signature of a hot chromosphere between 2 and 7\,R$_\star$ and concluded that the hot extended chromosphere detected in the UV must have a very small filling factor. Recently, however, \cite{ogorman_2017} directly measured the gas temperature at 1.3\,R$_\star$ to be $\sim$1000\,K below the photospheric effective temperature (T$_\textrm{eff}$ = 3690\,K), and $\sim$700\,K below the gas temperature at 2\,R$_\star$. This result indicates the decline of the gas temperature above the photosphere before it rises again toward higher temperatures. This trend in temperature is a characteristic of 1D semiempirical model atmospheres of cool evolved stars based on optical and UV diagnostics, including Betelgeuse \citep[e.g.,][]{basri_1981}.

The type II core-collapse supernova progenitor Antares ($\alpha$~Sco~A; M1.5 Iab), along with Betelgeuse, is the prototypical target for furthering our understanding of RSG chromospheres and mass loss. By applying our currently best-known values for stellar distances, it is possible that Antares is the nearest RSG ($d=170^{+35}_{-25}\,$pc, \citealt{van_Leeuwen_2007}), with Betelgeuse coming in a close second ($d=222^{+48}_{-34}\,$pc, \citealt{harper_2017}). Antares presents one of the largest photospheric angular diameters in the sky ($\phi_{LD} = 37.3\pm 0.1$ mas, \citealt{ohnaka_2013}), and unlike Betelgeuse, it has a wide binary companion ($\alpha$~Sco~B; B2.5~V) that is a hot main-sequence star located close to $2\arcsec.73$ west of the RSG \citep{reimers_2008}. This acts as an in situ probe of the RSG outer wind, making it a benchmark system for determining accurate mass-loss rates from RSGs \citep{kudritzki_1978}. The orbit is seen nearly edge on, with a period of $\sim2600$ years \citep{reimers_2008}. At radio wavelengths, \cite{hjellming_1983} found that the B2.5 V companion creates an optically thin H\,II region within the RSG wind. They constructed a model that provided a good representation of the observed ionization cavity, whose shape is an excellent diagnostic for the mass-loss rate of the RSG ($\dot{M} = 2\pm 0.5 \times 10^{-6}\,{\rm M}_\odot\,{\rm yr}^{-1}$, \citealt{braun_2012}). A subsequent study by \cite{brown_2004} showed that the radio emission from the RSG is spatially extended, but their data had insufficient sensitivity for detailed analysis. The order-of-magnitude increase in bandwidth now available for continuum observations offered by the Karl G. Jansky Very Large Array (VLA) enables a more sensitive study of the extended atmosphere of Antares. Moreover, the long baselines now offered by the Atacama Large Millimeter/submillimeter Array (ALMA) provide the possibility for a multiwavelength study of the extended atmosphere  of an RSG between $1-2$\,R$_\star$ for the first time. 


\section{Observations and data reduction}

\begin{table*}
\caption{VLA and ALMA observations of Antares.}
\label{tab1}
\centering
\begin{tabular}{c c c c c l c l c  }
\hline\hline
                                Date    & Band  & Bandwidth & Central & Time on   & Synthesized  &  rms  noise \\
                        &       &  & frequency & source   & beam FWHM  &   \\
                                        &       & (GHz) &(GHz) & (min)   & ($\arcsec\, \times\, \arcsec$, $^{\circ}$)   & ($\mu$Jy beam$^{-1}$)   \\
\hline
\rule{-2.6pt}{2.5ex} 2016 Oct 04 & S & 2& 3 &38 & $1.029 \times 0.417$, 12$^{\circ}$  &  10\\
                                         2015 Jun 18 & C & 4& 6 &49 & $0.695 \times 0.369$, 5$^{\circ}$ &  7\\
                     2015 Jun 21 & X & 4& 10 &25 & $0.425 \times 0.225$, 2$^{\circ}$ &  6\\
                     2015 Jun 21 & Ku & 6& 15 &18 & $0.231 \times 0.091$, 5$^{\circ}$ &  7\\
                     2015 Jun 21 & K & 8& 22 &20 & $0.157 \times 0.063$, 8$^{\circ}$ & 18\\
                                         2016 Dec 19 & Ka & 8& 33&41 & $0.101 \times 0.052$, 176$^{\circ}$ & 20\\
                     2016 Dec 09 & Q & 8& 44 &36 & $0.092 \times 0.051$, 176$^{\circ}$ & 95\\
                     2017 Oct 06  & 3 & 8& 97.5 & 5& $0.045 \times 0.043$, 146$^{\circ} $ & 53\\
                     2017 Sep 14 & 4 & 8& 145 & 5& $0.073 \times 0.039$, 44$^{\circ} $ & 86\\
                     2017 Jul 29 & 7 & 8& 343& 4& $0.075 \times 0.052$, 100$^{\circ} $ & 175\\
                     2017 Jul 29 & 8 & 8& 405 & 3& $0.063 \times 0.044$, 99$^{\circ}$ & 340\\
                                         
\hline
\end{tabular}
      \vspace{-2mm}
     \tablefoot{The synthesized beam dimensions and the rms noise values are taken from the Briggs weighted (robust = 0) images. }
\end{table*}

Antares was observed with ALMA in four unique continuum bands (bands 3, 4, 7, and 8) between July and October 2017, with 7.5\,GHz effective bandwidth spread over four spectral windows in each of the bands (project code: 2016.1.00234.S, PI: Eamon O'Gorman). A brief overview of these observations is presented in the lower part of Table \ref{tab1}. The observations in band 3 were centered on 97.5 GHz (3.1 mm) and were taken on 2017 October 6 using 45 antennas, with baselines ranging from 41 m to 16.2 km. The observations in band 4 were centered on 145 GHz (2.1 mm) and were taken on 2017 September 14 using 42 antennas, with baselines ranging from 41 m to 12.1 km. Finally, the observations in bands 7 and 8 were centered on 343 GHz (0.87 mm) and 405 GHz (0.74 mm) respectively, and were both taken on 2017 July 29 using 47 antennas, with baselines ranging from 17 m to 3.7 km. The total observing time for each of the four tracks was approximately 20 minutes in duration with only between 3 and 5 minutes spent on the target in each track. The quasar J1625-2527, which was within 2$^\circ$ of Antares, was used as the gain calibrator for all four bands. The quasar J1617-2537 was used as the secondary gain calibrator in band 3, while the quasar J1626-2951 was used in bands 4, 7, and 8. The quasar J1517-2422 from the ALMA quasar catalog served as both bandpass and absolute flux density calibrator for all ALMA observations. The uncertainty on this flux density standard is expected to be less than 10\% in bands 3, 4, and 7 and less than 15\% in band 8 \citep{fonalont_2014}. 

\begin{figure*}[hbt!]
\centering 
\mbox{
\includegraphics[trim=0pt 0pt 20pt 0pt,clip, scale=0.28]{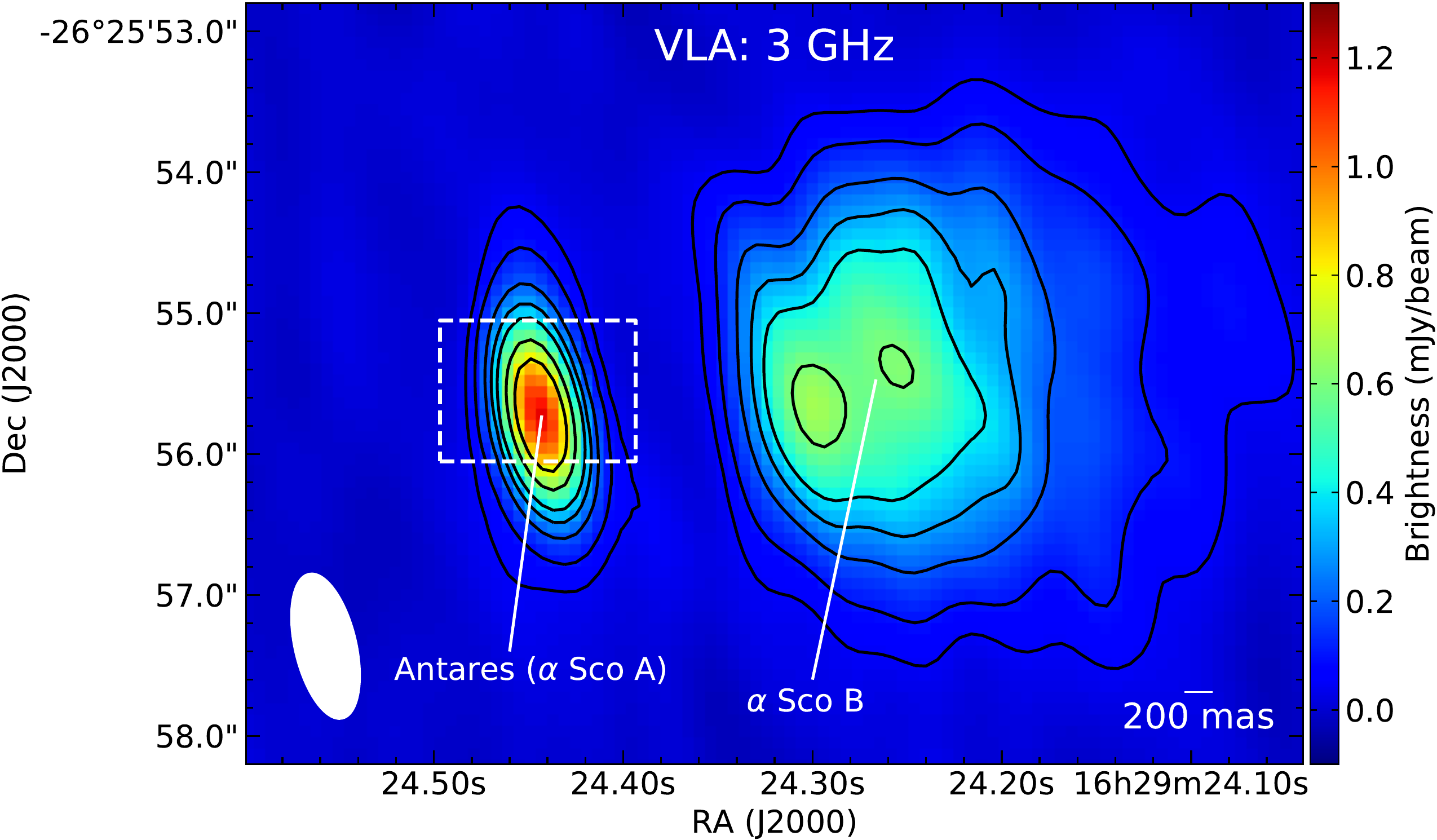}
\includegraphics[trim=123pt 0pt 20pt 0pt,clip, scale=0.28]{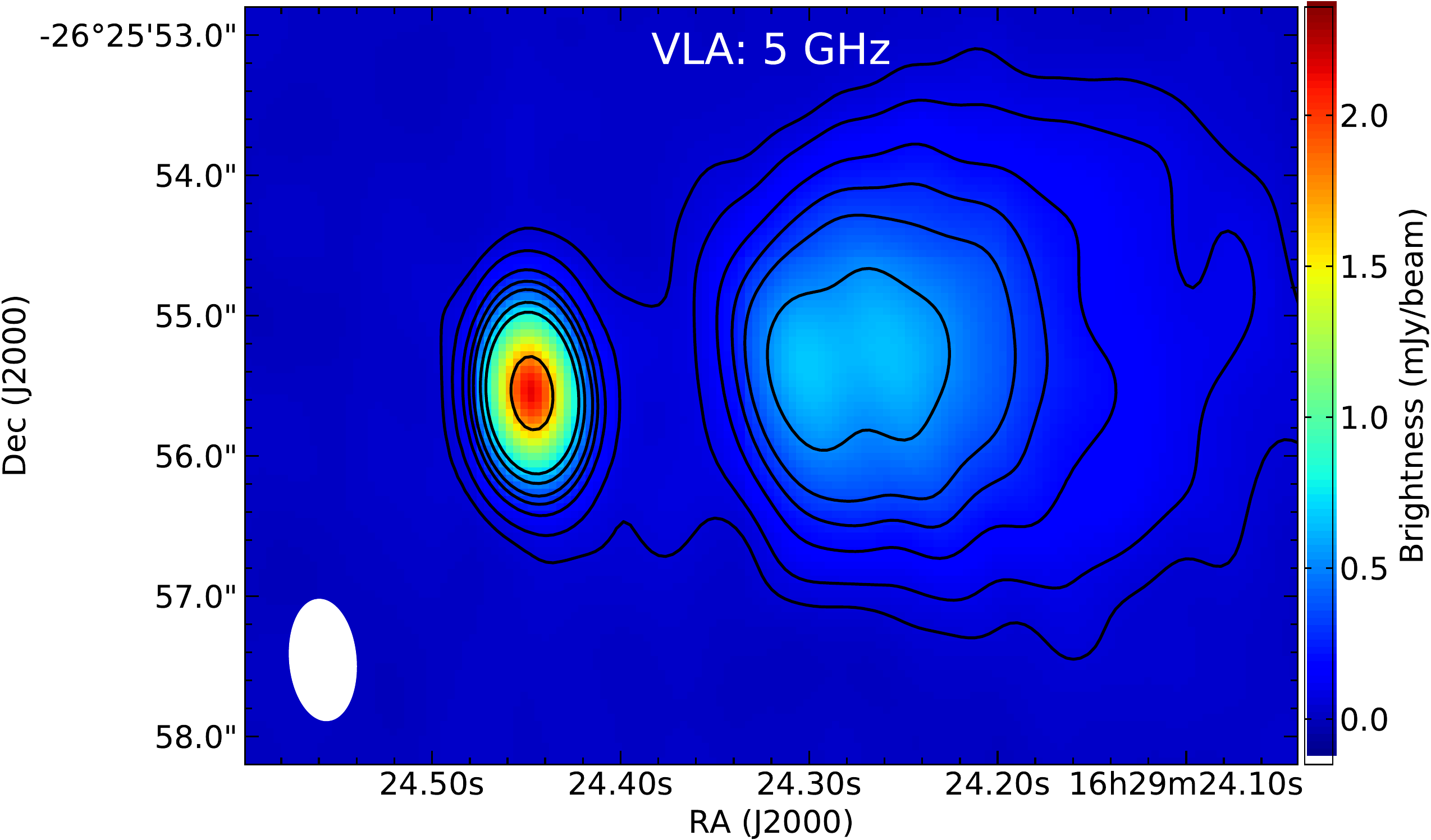}
\includegraphics[trim=123pt 0pt 0pt 0pt,clip, scale=0.28]{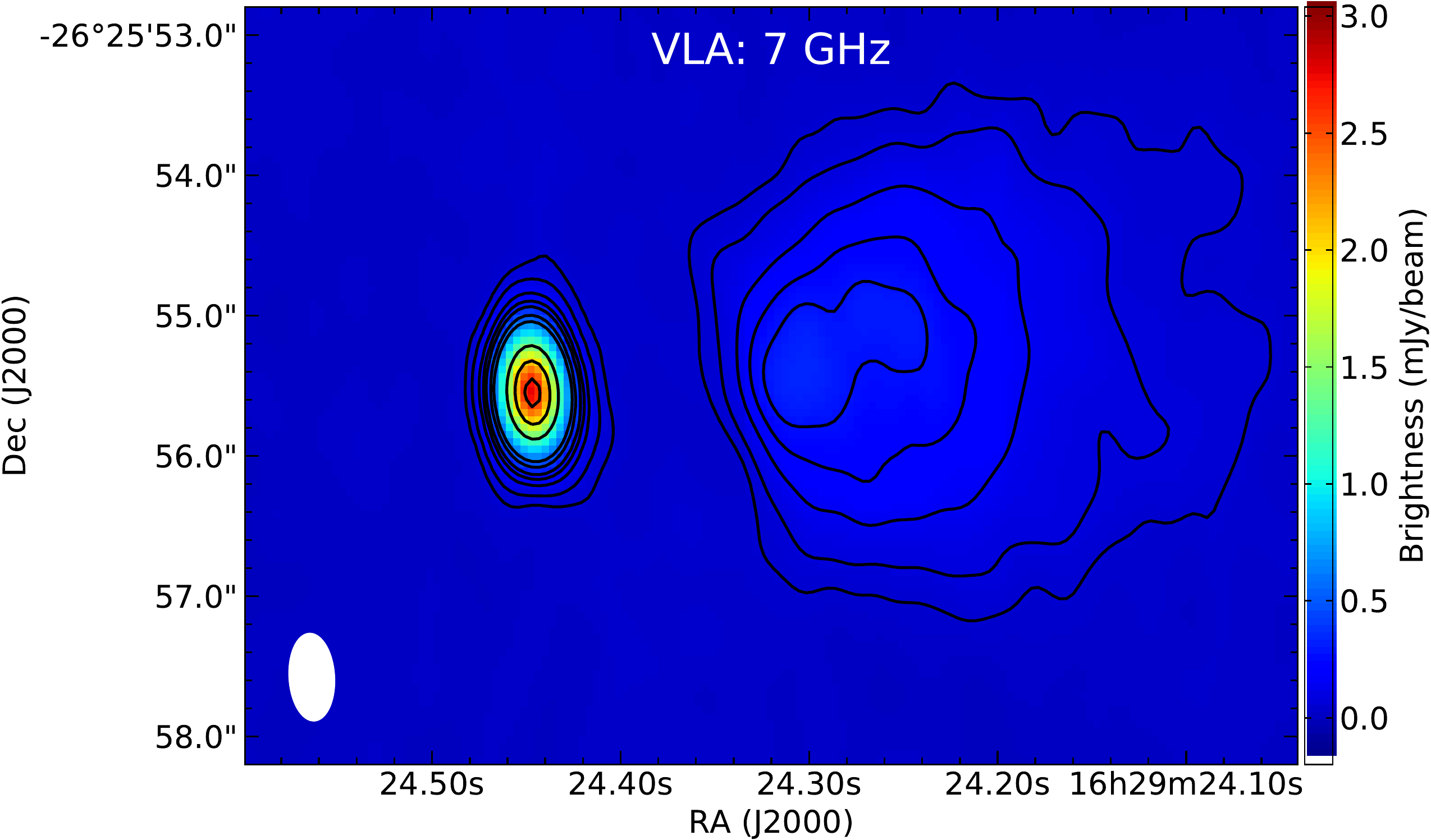}
}
\mbox{
\includegraphics[trim=0pt 0pt 20pt 0pt,clip, scale=0.28]{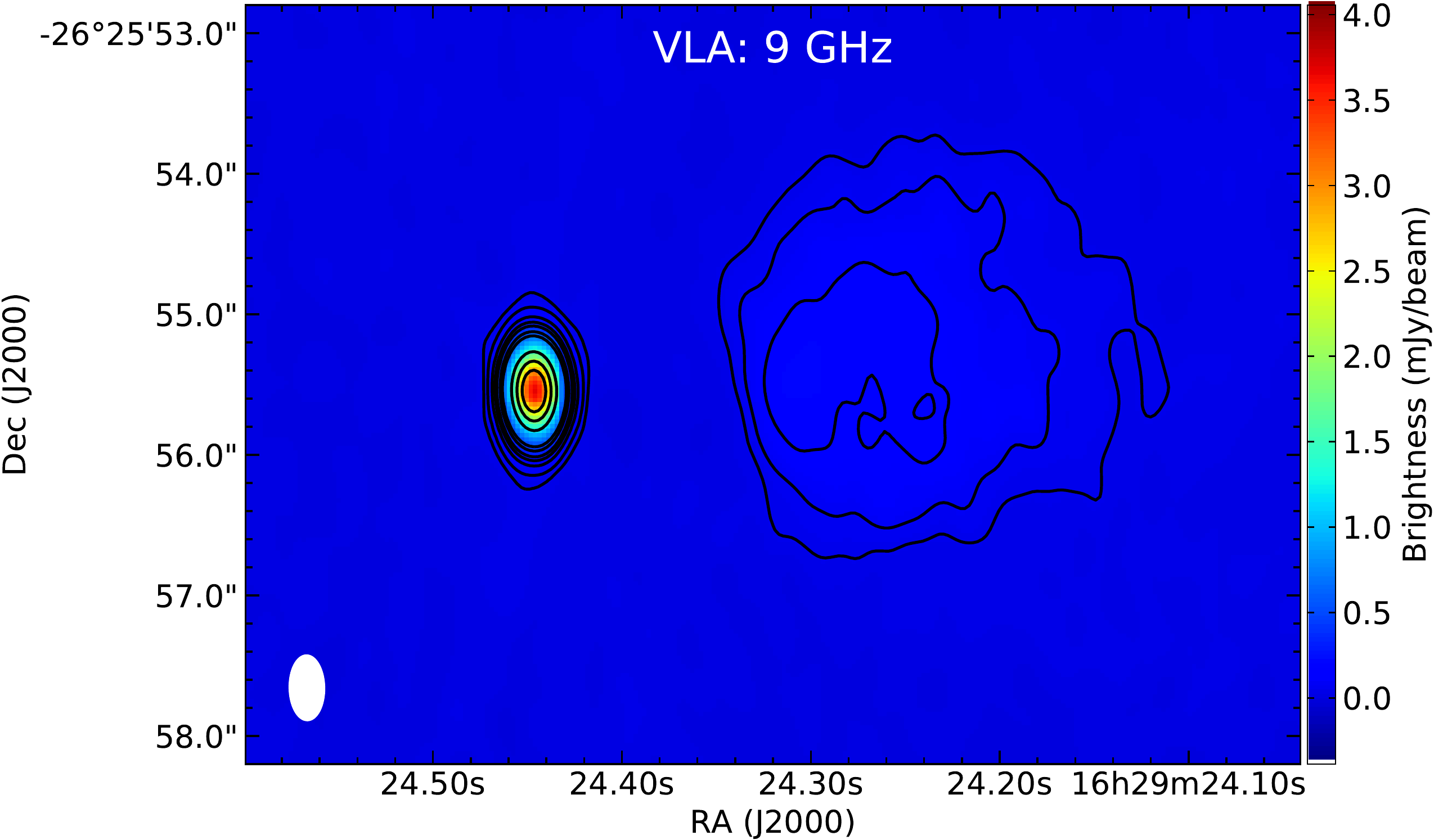}
\includegraphics[trim=123pt 0pt 20pt 0pt,clip, scale=0.28]{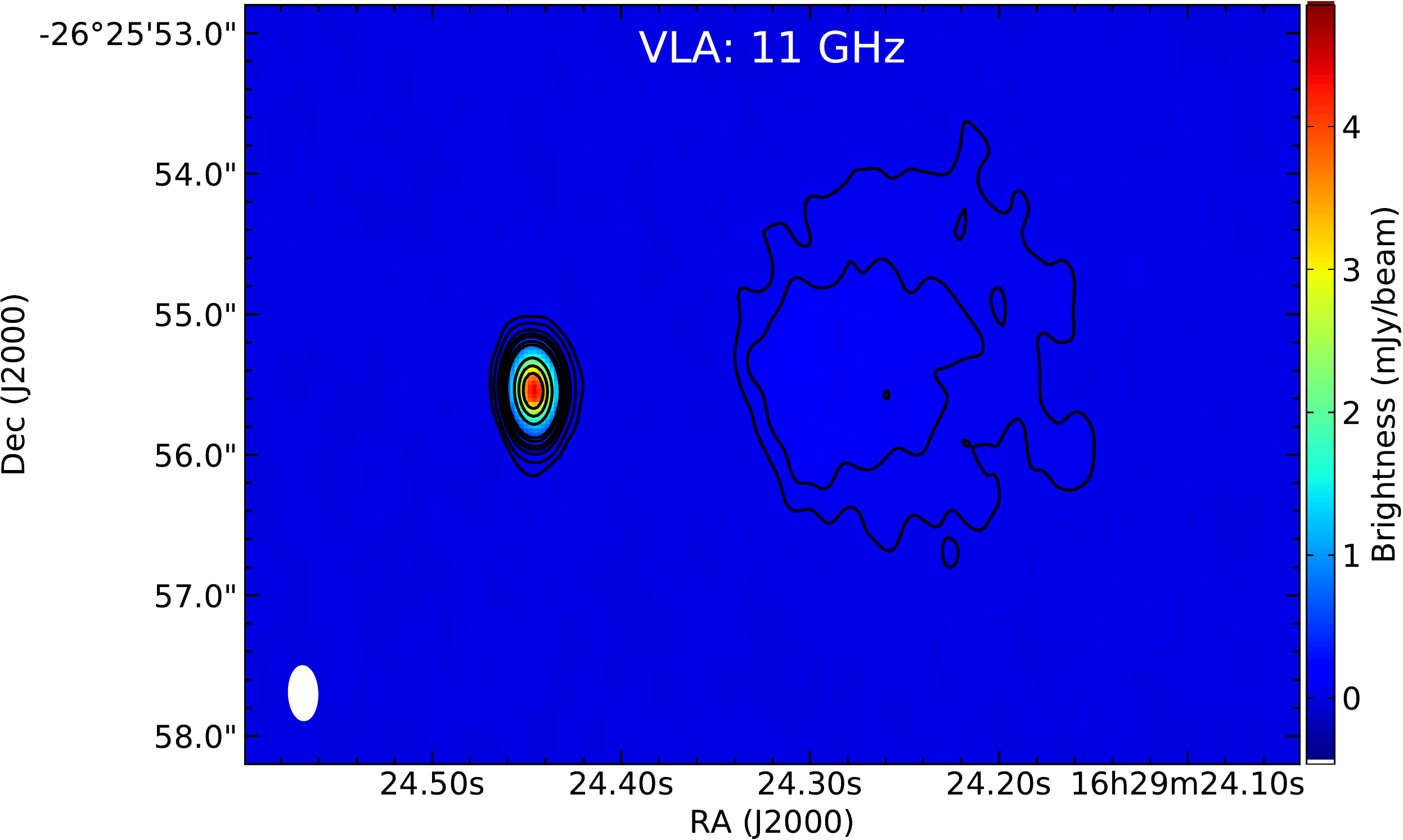}
\includegraphics[trim=123pt 0pt 0pt 0pt,clip, scale=0.28]{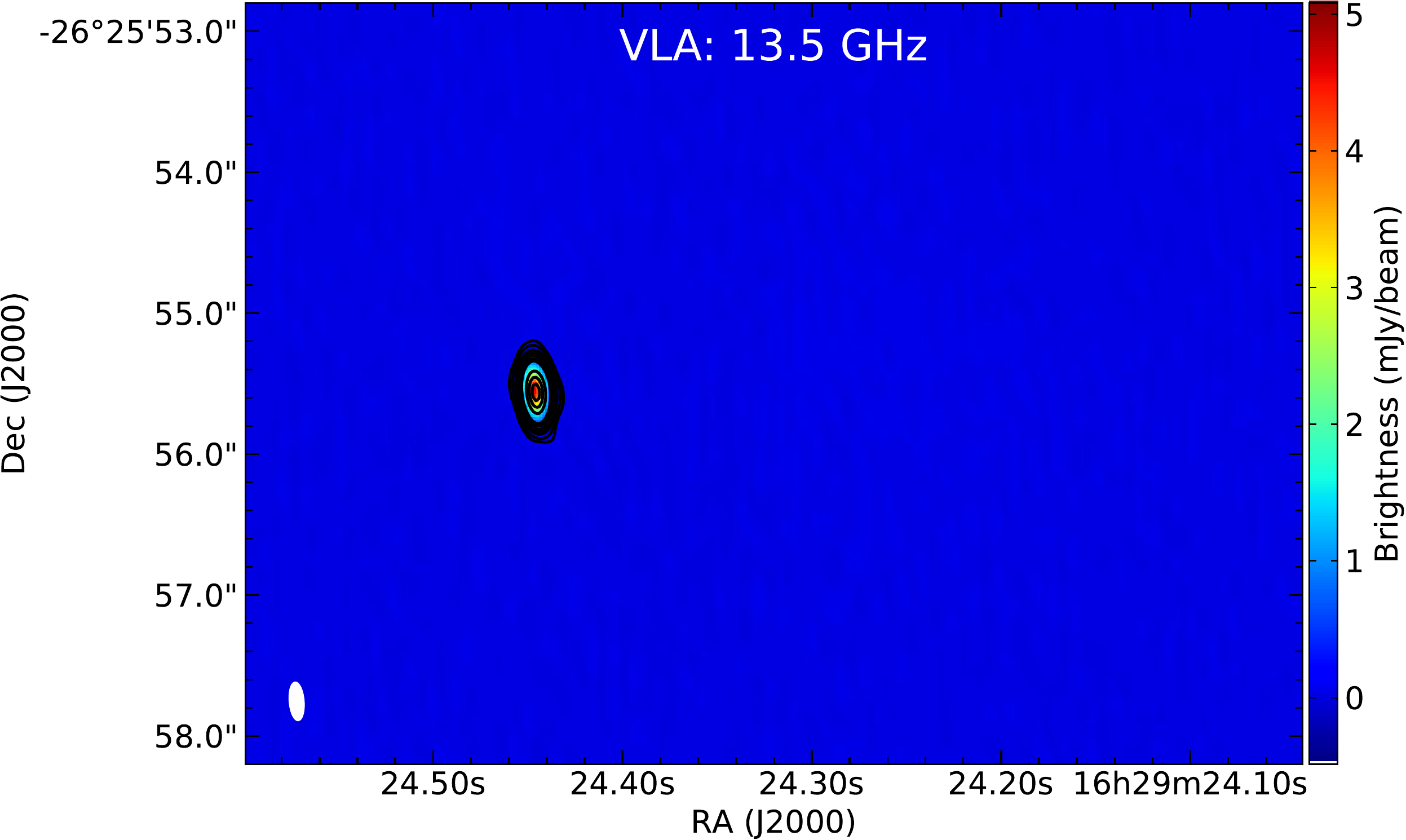}
}
\mbox{
\includegraphics[trim=0pt 0pt 20pt 0pt,clip, scale=0.28]{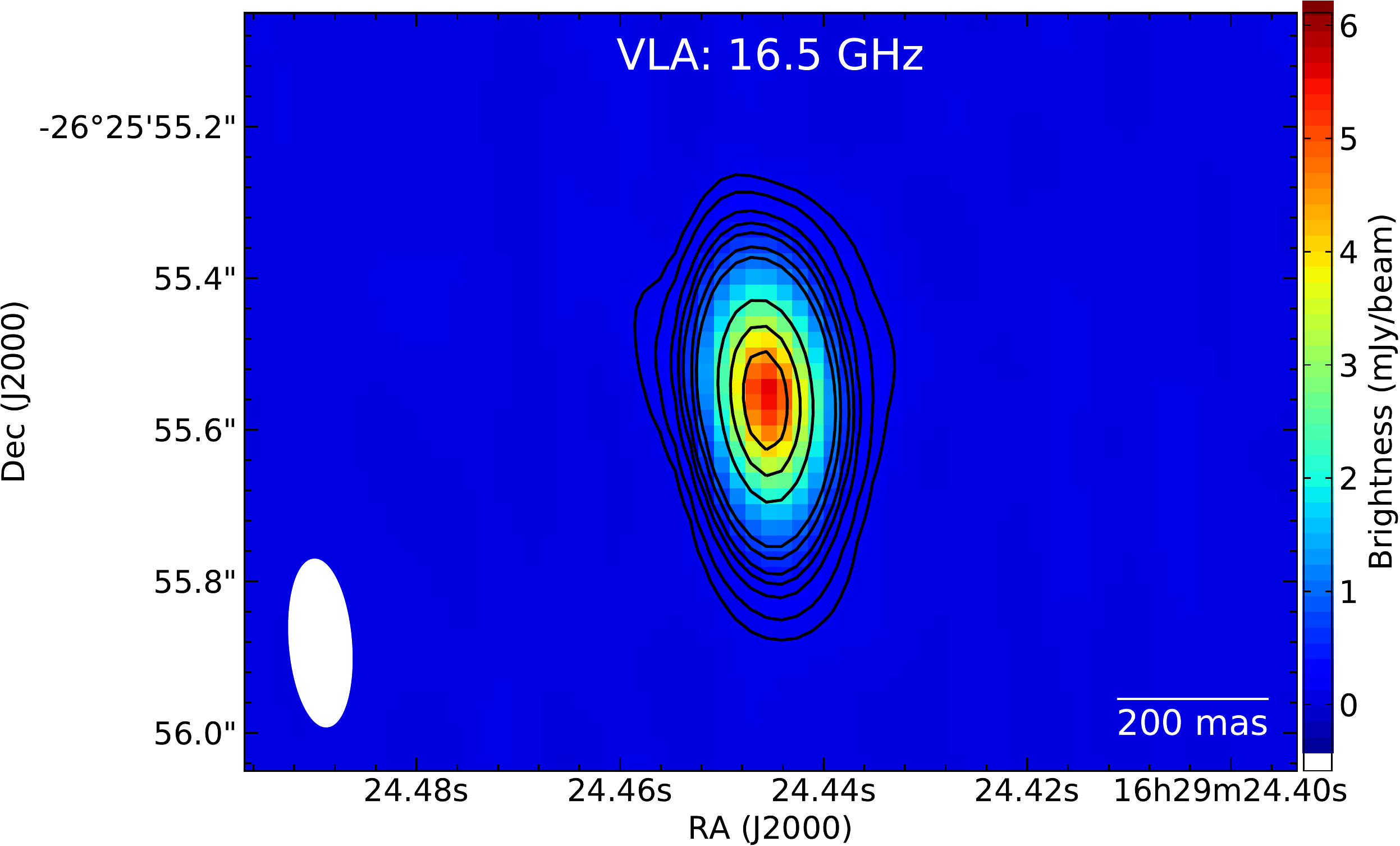}
\includegraphics[trim=123pt 0pt 20pt 0pt,clip, scale=0.28]{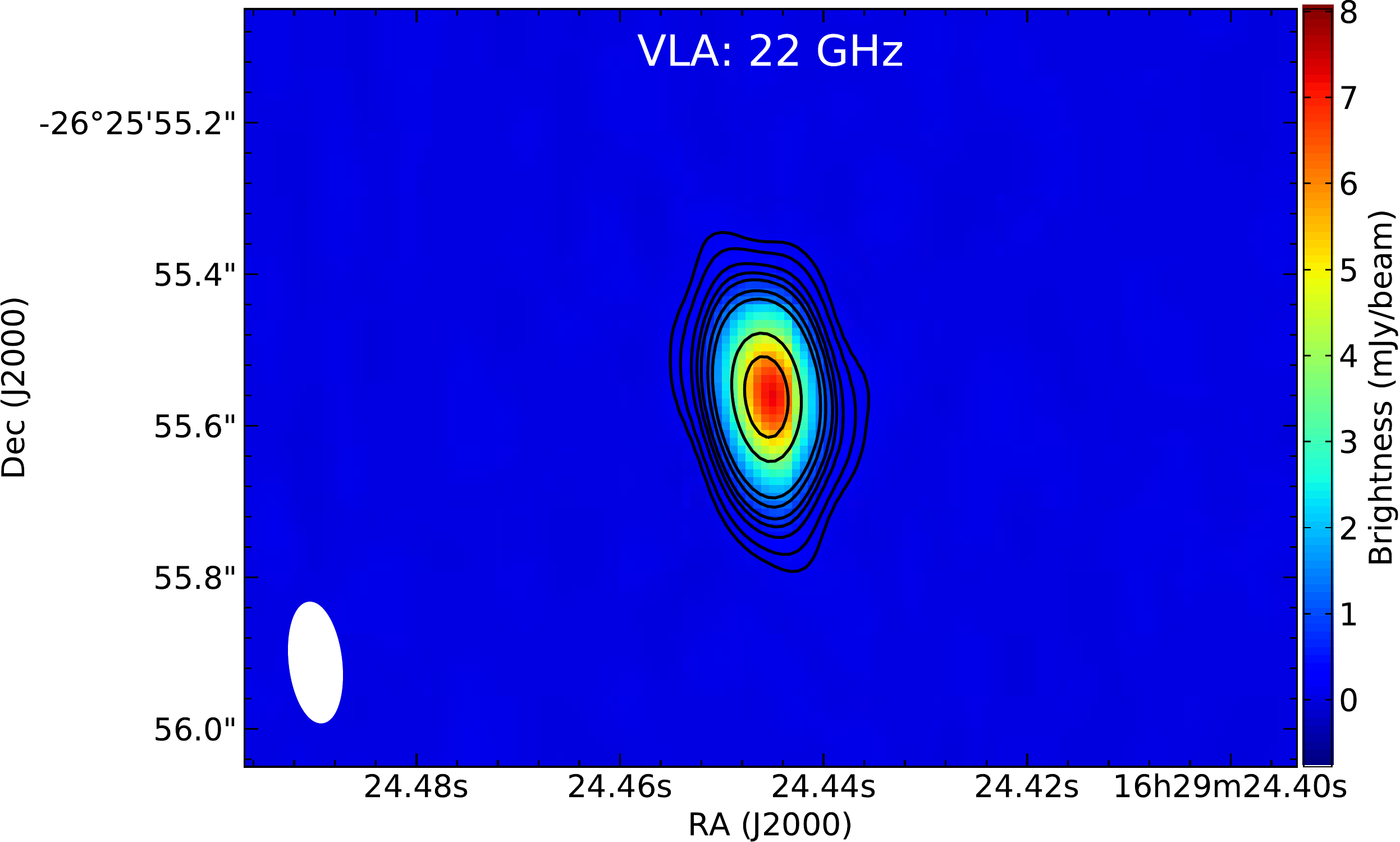}
\includegraphics[trim=123pt 0pt 0pt 0pt,clip, scale=0.28]{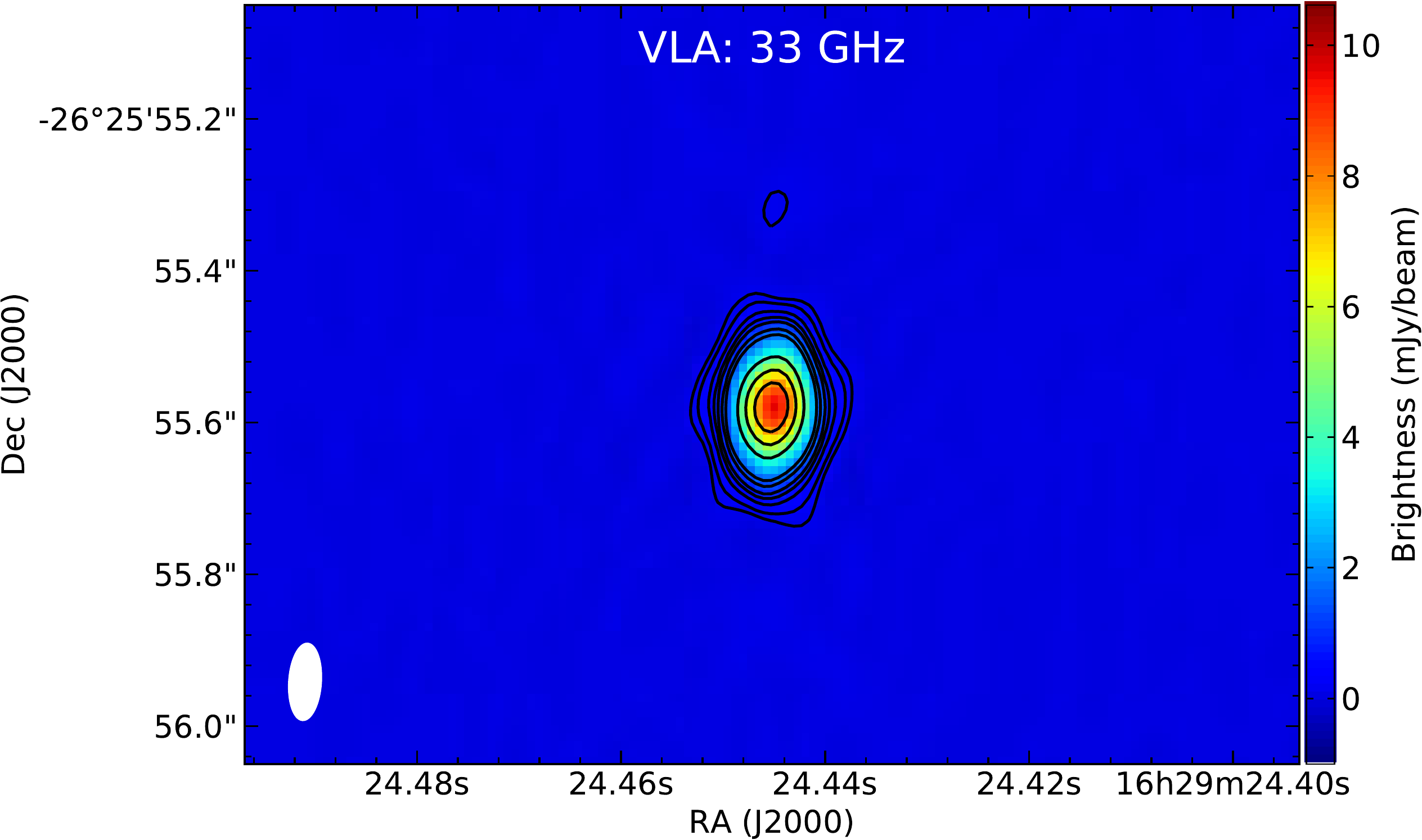}
}
\mbox{
\includegraphics[trim=0pt 0pt 20pt 0pt,clip, scale=0.28]{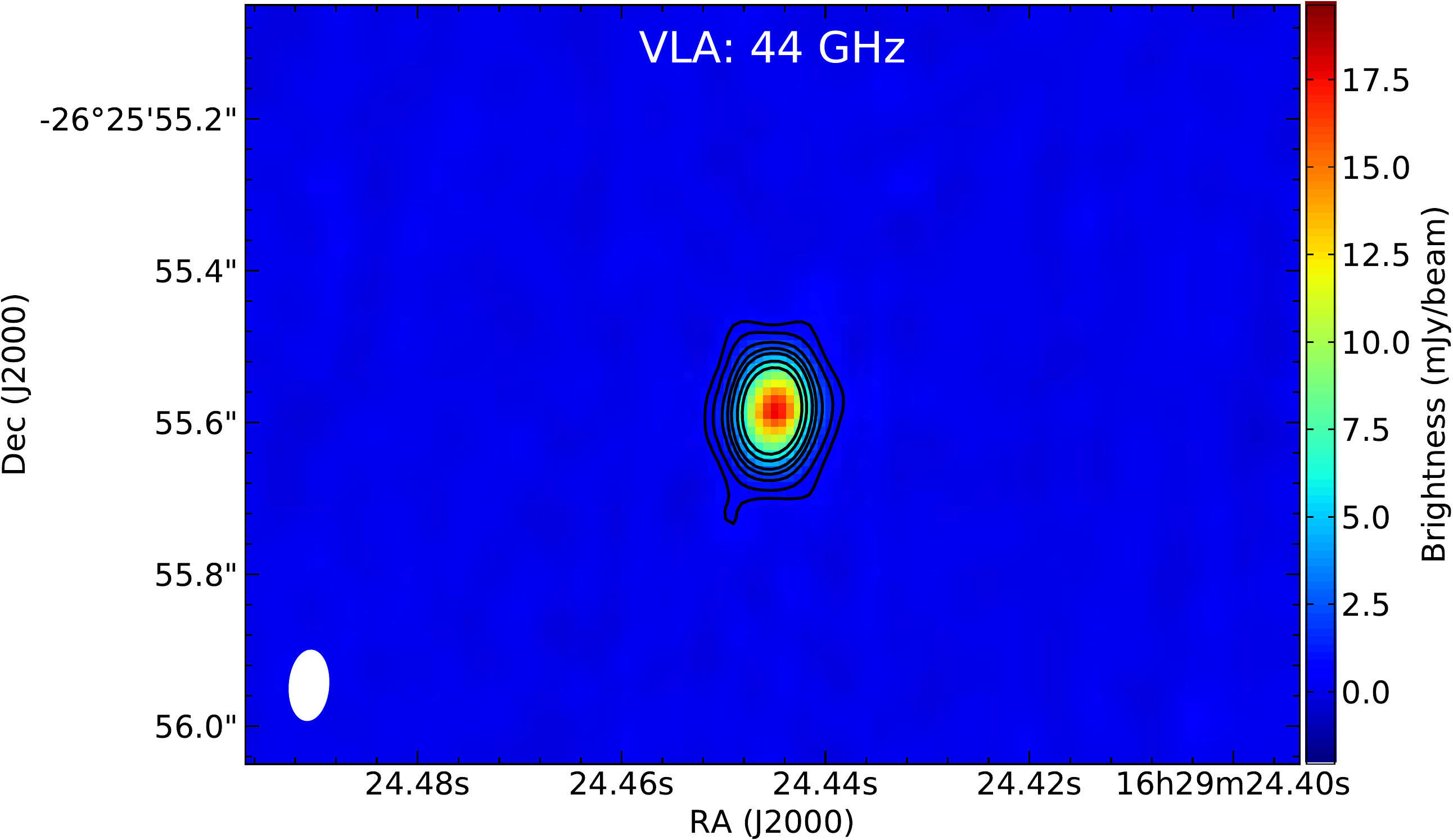}
\includegraphics[trim=123pt 0pt 20pt 0pt,clip, scale=0.28]{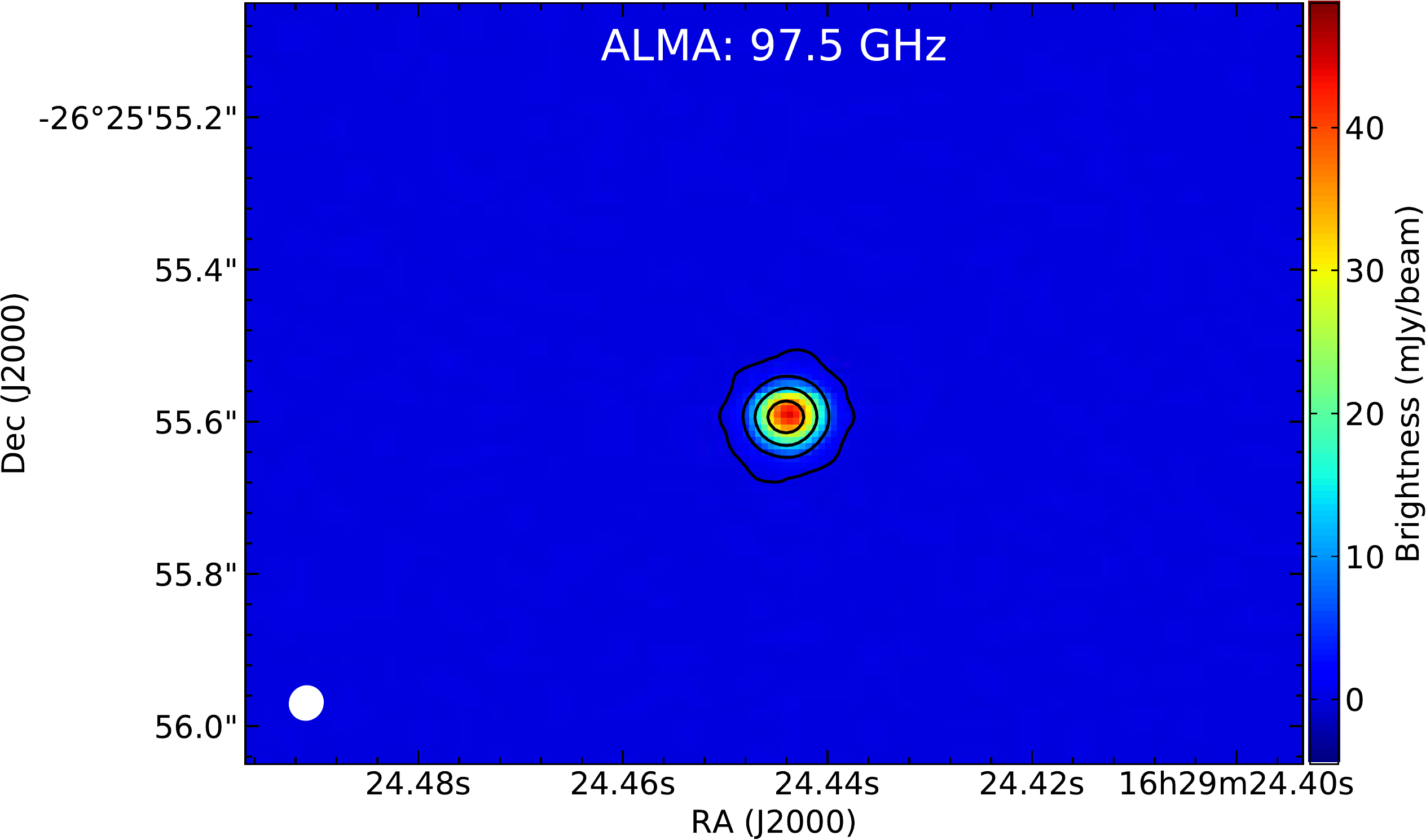}
\includegraphics[trim=123pt 0pt 0pt 0pt,clip, scale=0.28]{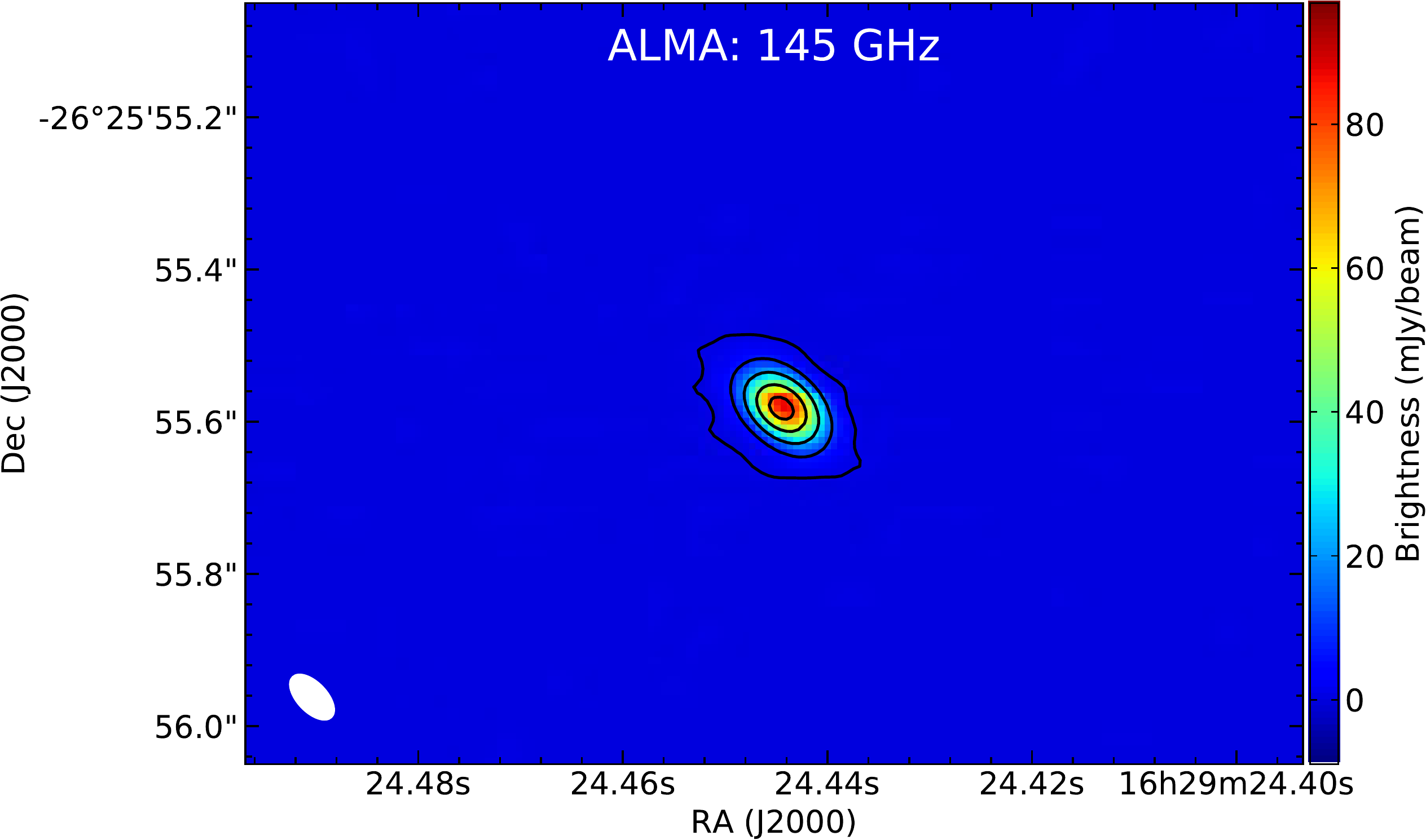}
}
\mbox{
\includegraphics[trim=0pt 0pt 20pt 0pt,clip, scale=0.28]{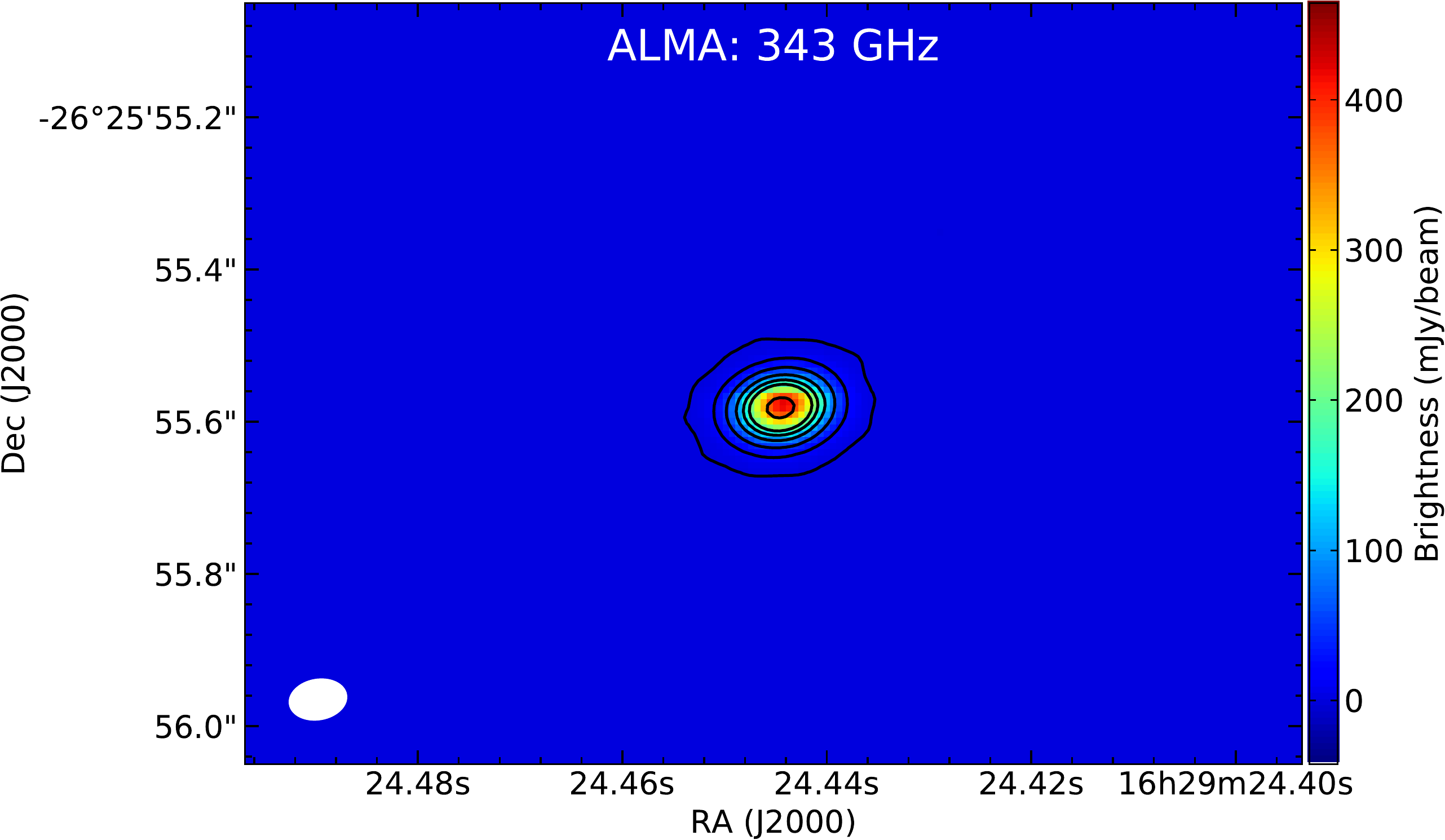}
\includegraphics[trim=123pt 0pt 0pt 0pt,clip, scale=0.28]{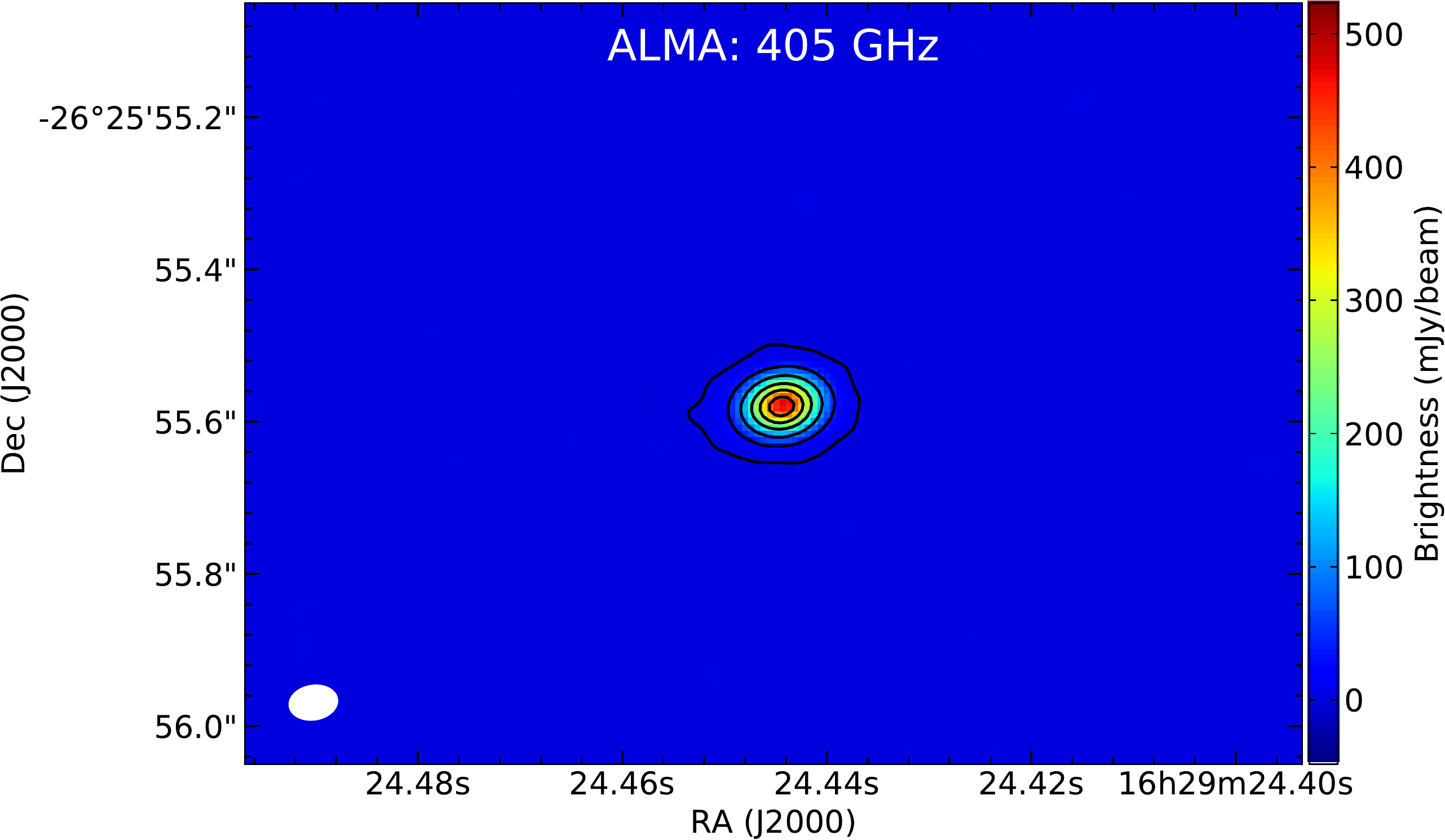}
}
\caption[]{VLA and ALMA Briggs-weighted (robust 0) images of Antares at frequencies between 3 and 405\,GHz. The optically thin  emission from the ionized cavity in the outer wind of Antares is caused by the UV radiation field of its hot B-type companion and is not detected in our images above 11\,GHz. The size and shape of the restoring beam at each frequency is shown in the lower left corner of each image. Contour levels are set to $(5, 10, 20, 30, 40, 60, 80, 200, 300, 400)\times \sigma _{\textrm{rms}}$ and $(5, 100, 300, 600, 900, 1200, 2100)\times \sigma _{\textrm{rms}}$ for the VLA and ALMA images, respectively. The dashed box in panel 1 indicates the region depicted in panels 7 -14.}
\label{fig1}
\end{figure*}

Antares was observed with the VLA in the most extended A configuration in all frequency bands between 2 and 50 GHz (bands S, C, X, Ku, K, Ka, and Q) between June 2015 and December 2016 (project codes: 15A-054 and 16B-012, PI: Eamon O'Gorman). An overview of these observations is presented in Table \ref{tab1}. The total observing time of Antares per band ranged from 18 minutes to 49 minutes. Each individual target scan was interleaved between scans of a gain calibrator, ranging in duration from 0.5 minutes at 45 GHz (Q band) to 1 minute at 3 GHz (S band). Like the ALMA observations, the quasar J1626-2527 was also used as the gain calibrator at all frequencies, except at S band, where the quasar J1626-2951 (located within 4$^\circ$ of Antares) was used instead. 3C286 was used to calibrate the absolute flux density scale and instrumental bandpass at all VLA bands. The uncertainties in the absolute flux densities of the VLA observations are conservatively assumed to be 5\% at S through Ku bands and 15\% for the K, Ka, and Q bands\footnote{https://science.nrao.edu/facilities/vla/docs/manuals/oss/performance/fdscale}.

The ALMA and VLA data were calibrated using standard ALMA and VLA calibration pipelines and manually inspected and imaged using CASA 4.7.2. Stokes I images were produced using the CLEAN task, with the multifrequency synthesis mode, two Taylor coefficients to model the frequency dependence, and Briggs weighting with a robust parameter of 0. The sizes of the synthesized beams and the rms sensitivity of the images are given in Table \ref{tab1}. Antares was sufficiently bright at all wavelengths to allow for two to three rounds of phase self-calibration and one round of amplitude self-calibration. The basic properties of Antares such as flux density, shape, and size are derived in Section \ref{sec3}, and to do so, we analyzed the \textit{uv}-data directly using the UVMULTIFIT code \citep{vidal_2014}. At the long VLA wavelengths (S, C, and X bands), the visibilities also contained strong emission from the ionized region in the Antares wind due to its hot companion. This emission needed to be subtracted from the visibilities before accurate \textit{uv}-fitting could be carried out. To do so, a model of the emission from the ionized region alone was created using the CLEAN task. Using the FT task, this model was then Fourier transformed to produce a model of the visibilities for the ionized cavity. Finally, the UVSUB task was used to subtract this model from the actual visibilities to leave a dataset containing only the visibilities of Antares. 

\section{Results}\label{sec3}
\begin{table*}
\caption{Parameters from the uniform circular disk fits to the VLA calibrated visibilities and brightness temperature values.}
\label{tab2}
\centering
\begin{tabular}{c c c c c c c c c  }
\hline\hline
                                Frequency       & Wavelength    & Flux density & Spectral index & Modeled diameter   & Brightness temperature \\
                        (GHz)   & (cm)  & (mJy) &  & (mas)   & (K)     \\
\hline
\rule{-2.6pt}{2.5ex} 3 & 10.0 & $1.56\pm0.01$ & $0.86\pm 0.03$ & $430.7\pm9.1$ & $1647\pm 110$ \\
                                         5 & 6.0 & $2.34\pm 0.01$ & $0.86\pm0.04$ & $298.9\pm 5.2$& $1849\pm 113$ \\
                     7 & 4.3 & $3.07 \pm 0.01$ & $0.94\pm0.03$ & $223.2\pm 2.2$& $2215\pm 120$ \\
                     9 & 3.3 & $4.10\pm 0.01$ & $1.08\pm0.03$ & $177.6\pm 1.7$& $2825\pm 152$ \\
                     11 & 2.7 & $4.97\pm 0.02$& $1.10\pm 0.04$ & $155.4\pm 1.3$& $2997\pm 156$ \\
                     13.5 & 2.2 & $6.10\pm 0.02 $& $1.35\pm 0.02$ & $128.9\pm 0.6$& $3546\pm 181$ \\
                     16.5 & 1.7 & $7.74\pm 0.02 $& $1.26\pm 0.03$ & $117.8\pm 0.4$ & $3609 \pm 183$ \\
                     22 & 1.4 & $10.57\pm 0.02$ & $1.31\pm 0.01$ & $100.9\pm 0.1$& $3777\pm 567$ \\
                                         33 & 0.9 & $19.59 \pm 0.02$& $1.05\pm 0.01$& $92.1\pm 0.1$& $3736 \pm 560$ \\
                     44 & 0.7 & $26.86 \pm 0.10$& $1.44\pm 0.06$ & $80.5\pm 0.4$& $3772\pm 567$ \\
                                         
\hline
\end{tabular}
      \vspace{-2mm}
     \tablefoot{The errors on the flux density, spectral index, and diameter are the fitting errors. The errors on the brightness temperature contain the systematic error on the flux density.}
\end{table*}

\begin{table*}
\caption{Parameters from the uniform elliptical disk fits to the ALMA calibrated visibilities and brightness temperature values.}
\label{tab3}
\centering
\begin{tabular}{c c c c c c c c c c c }
\hline\hline
                                Frequency       & Wavelength    & Flux density & Spectral index & Major axis   & Axis ratio & P.A. &Brightness temperature \\
                        (GHz)   & (cm)  & (mJy) &  & (mas)  & & $^{\circ}$ & (K)     \\
\hline
\rule{-2.6pt}{2.5ex} 97.5 & 0.31 & $90.05\pm0.04$ & $1.28\pm 0.01 $ & $74.2\pm0.1$ & $0.85\pm0.01$ & $93.9\pm0.2$& $3566\pm 357$\\
                                         145 & 0.21 & $146.18\pm0.03$ & $1.27\pm0.01$ & $67.9\pm0.1$& $0.85\pm0.01$ & $95.1\pm0.2$ & $3125\pm 313$\\
                     343 & 0.09 & $530.03\pm0.08$ & $1.74\pm0.01$ & $56.3\pm0.1$& $0.86\pm0.01$ &$95.8\pm0.2$& $2911\pm 291$\\
                     405 & 0.07 & $642.57 \pm 0.20$ & $1.59\pm0.02$ & $54.7\pm0.1$& $0.86\pm0.01$ & $96.5\pm0.3$& $2682\pm 402$\\
                                         
\hline
\end{tabular}
      \vspace{-2mm}
     \tablefoot{All errors shown are the fitting errors, except for the errors on the brightness temperature, which also include the systematic error on the flux density.}
\end{table*}
In Figure \ref{fig1} the ALMA and VLA images of the target system are shown. Simple Gaussian fits to Antares confirm that the star is spatially resolved at all ALMA and VLA wavelengths. However, with the synthesized beam being generally of similar size to the star at all wavelengths, no properties of Antares were derived from these images and were instead derived from the visibilities. The VLA images at the longest wavelengths are the most sensitive to low surface brightness emission. These images clearly show the extended ionized region of the circumstellar atmosphere of Antares that is illuminated by the UV field of its hot B spectral type companion. The VLA A configuration is sensitive to emission on spatial scales smaller than 18$\arcsec$, 9$\arcsec$, and 5$\arcsec$ at S, C, and X band. We found no evidence for the existence of emission on similar or larger spatial scales that could have been resolved out by the VLA.  Although not the focus of this paper, the morphology of this H\,II region is fully consistent with that presented by \cite{newell_1982}. The superior dynamic range of our images confirm the double-lobe structure modeled previously \citep{braun_2012} and also tentatively detected by \cite{newell_1982}. Moreover, the large dynamic range of our images also allow spectral index images to be created from within each band. The long VLA wavelength spectral index images confirm that the emission from the entire ionized region of the circumstellar atmosphere is consistent with being optically thin, that is, the spectral index values are consistently close to $\nu = -0.1$, again in agreement with the findings of \cite{newell_1982}. We emphasize that none of the emission from the H\,II region is included in any of the analysis in the subsequent sections.

\begin{figure}[ht!]
\centering 
\includegraphics[angle=90, scale=0.41]{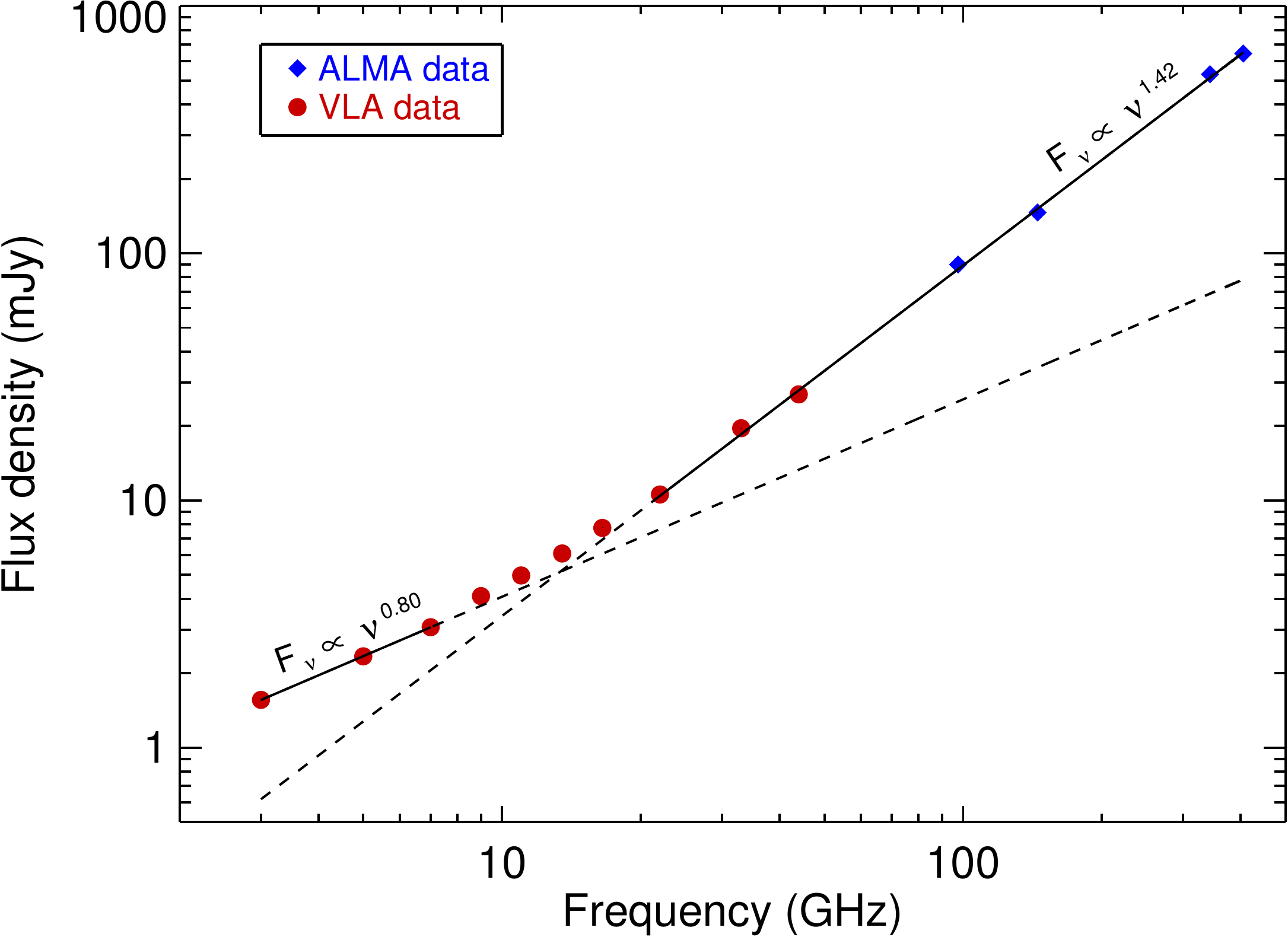}
\caption[]{Radio spectral energy distribution for the red supergiant Antares. The filled red circles and the filled blue diamonds represent the VLA and ALMA data, respectively, and include the statistical 1$\sigma$ error bars. The continuous lines represents the best-fit line to the underlying data points, while the dashed lines are extrapolations to the remaining data points. The slope on the lines (i.e., the spectral index values) clearly changes from 1.42 above 20 GHz to 0.80 below 8\,GHz.}
\label{fig2}
\end{figure}

\begin{figure}[ht!]
\centering 
\includegraphics[angle=90, scale=0.41]{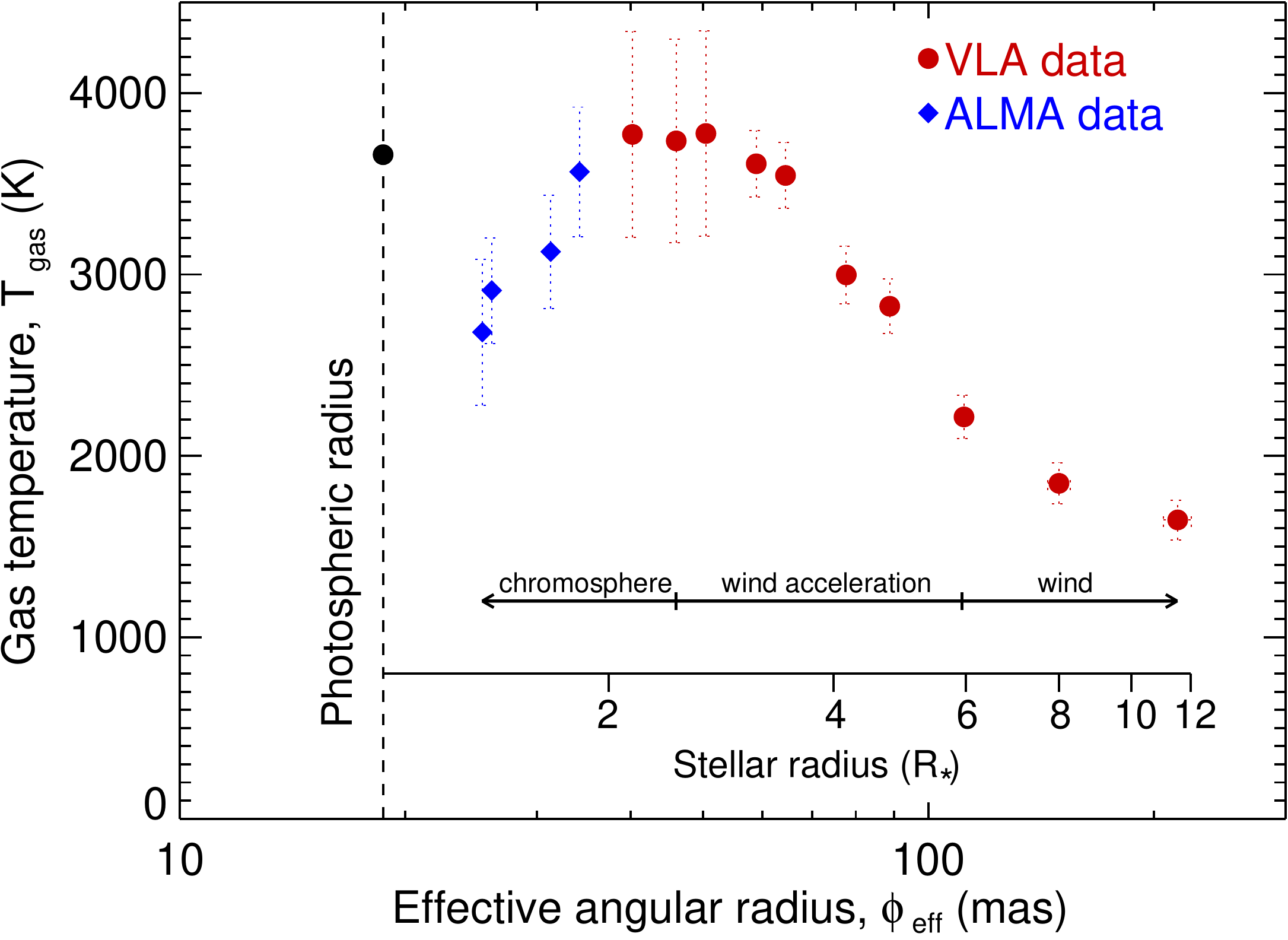}
\caption[]{Empirically derived gas temperature of the Antares atmosphere as a function of distance from the optical photosphere and projected radius. The filled blue diamonds are the values derived from the ALMA data, while the filled red circles are the values derived from the VLA data. The error bars on the gas temperature include the absolute flux density uncertainty and the fitting errors. The filled black circle at R$_{\star}$=1 represents the photospheric effective temperature of 3660\,K for Antares \citep{ohnaka_2013}.}
\label{fig3}
\end{figure}

\subsection{Size and shape of the Antares atmosphere}

A number of relatively simple models were fit to the ALMA and VLA visibilities of Antares. The models ranged from uniform intensity circular and elliptical disks to slightly more complex circular and elliptical disks with one and two superimposed point sources, Gaussians and rings; all of which have been used to model radio emission from other evolved stars \citep[e.g.,][]{lim_1998,ogorman_2017, matthews_2018}. Uniform-intensity circular disks provided the best fits to all of the VLA data, although this could be due to the fact that the star is only marginally resolved at most VLA wavelengths, but it is better resolved with ALMA, which provides better distinction between simple and more complex models. The spectral index within each band was included as another free parameter to these circular disk fits due to the large fractional bandwidth of our VLA observations. While this did not improve the fits further, neither did it change the values of the other parameters significantly. The parameters of these fits are given in Table \ref{tab2}. We note that although the spectral index value within each band may not be precise, there is generally a trend to higher values at higher VLA frequencies. The other interesting property from these circular fits is that the diameter of Antares continuously decreases from a vast size of 431 mas (23 R$_{\star}$) at 3 GHz to just 80.5 mas (4.3 R$_{\star}$) at 44 GHz. This trend in size with frequency is expected from any optically thick stellar atmosphere where the radio opacity is dominated by thermal free-free interactions \citep{wright_1975, panagia_1975}. 

Uniform-intensity elliptical disk models in which the flux density depended on frequency (i.e., a spectral index) were found to fit all of the four ALMA datasets best. The residual images whereby these best-fit uniform-intensity elliptical disk models were subtracted from the data are briefly discussed in Appendix \ref{fig1a}. The best-fit uniform-intensity elliptical disk models to the ALMA data are described in Table \ref{tab3}. We note that the errors in the diameter measurements are small and would only be precise if the atmosphere corresponded exactly to our chosen model. However, because no substantial residuals are detected in any of the residual images, we are confident that our model represents the data well. Again, like in the VLA data, there was a general increase in spectral index from low to higher values. Moreover, the major axis of the ellipse and the geometric mean of major and minor axes both decreased continuously as frequency increased. When we take this geometric mean as our definition for the stellar diameter at these frequencies, the star deceases in diameter from 68.4 mas (3.7 R$_{\star}$) at 97.5 GHz to 50.7 mas (2.7 R$_{\star}$) at 405 GHz. A notable feature of these elliptical models is that both the axis ratio (defined as the ratio of the minor axis to the major axis) and the position angle (P.A.; defined as angles east of north) are almost identical at all four frequencies, indicating that a large-scale asymmetry exists throughout at least this region of the atmosphere. The P.A. of the major axis is approximately aligned with the direction of the Antares companion, $\alpha$ Sco B. However, \cite{ogorman_2017} also found a similar global asymmetry for Betelgeuse, which is a single RSG. It might therefore well be that this asymmetry is intrinsic to the RSG itself and not connected to the presence of its companion. 

\cite{de_koter_1988} cited a value of $v\sin i$ for Antares of 10 km\,s$^{-1}$ from \cite{bernacca_1973}. If the RSG is rotating in the same sense as the orbit then $\sin i \approx 1$, and for reasonable estimates for the mass of Antares, 
this does not lead to any significant rotational distortion of the photosphere ($\sim 1\%$). In a photospheric abundance study, \cite{harris_1984} found a macroscopic broadening parameter, which is indicative of radial-tangential and rotational broadening, of
of 6 km\,s$^{-1}$ (Betelgeuse = 7 km\,s$^{-1}$). This means that there is no evidence of a large $v\sin i$ for Antares.

The elongation seen in the ALMA fits may indicate an equatorial enhancement of atmospheric heating, assuming the orbital and rotation axis are aligned.  This would suggest that it is not connected to a large convection cell origin and perhaps it might be related to the atmospheric magnetic field geometry.

\subsection{Radio spectral index of Antares} \label{sec3.2}
The ALMA and VLA flux densities of Antares derived from the \textit{uv}-fitting are plotted against frequency in Figure \ref{fig2}. It is immediately apparent that the spectral index varies across all of the sampled frequencies. We find that below 8 GHz the spectral index value is 0.8, while above 20 GHz, the value rises to 1.42. Between these two frequencies, the spectral index turns over to the two different values. This behavior is consistent with the trend in spectral index values listed in Tables \ref{tab2} and \ref{tab3}, in that they generally increase in values from low to high frequencies. \cite{newell_1982} derived a spectral index value of 1.05 for their three sampled frequencies (1.5, 4.9, and 15 GHz). From our work, it now appears that they sampled regions of the atmosphere where the value of the spectral index both remains constant and turns over. The greater sensitivity and much finer sampling of frequency of our ALMA and VLA data now show that in fact the spectral index value is higher than the original value of 1.05 above 20 GHz and lower below 10 GHz.

The change in the value of the spectral index across frequencies for Antares provides powerful diagnostic information about the different atmospheric layers that our ALMA and VLA data sample. For an isothermal source with an angular diameter independent of frequency (i.e., a hard disk), the spectral index should just reflect the frequency factor in the Planck function, that is, $\alpha = 2$. The spectral index for a spherically symmetric optically thick stellar wind with a constant wind velocity, temperature, and ionization fraction should be 0.6 \citep{wright_1975, panagia_1975, olnon_1975}. These two cases are generally an oversimplified view of a stellar atmosphere, and in reality, the spectral index values derived will be somewhere between the two. Our derived value of $\alpha = 0.8$ below 10 GHz is close to 0.6, which indicates that these wavelengths likely sample emission from the wind, although the reality of, for example, a wind temperature gradient and wind acceleration can explain the slight difference between the two values. Likewise, the higher value of 1.42 above 20\,GHz indicates that the sampled emission comes from regions closer to the stellar photosphere where the ionized density scale height is small compared to the radius, and the source begins to resemble a blackbody disk, but with a depth-dependent temperature.

\subsection{Temperature profile of the atmosphere}
The brightness temperatures at each ALMA and VLA frequency are listed in Tables \ref{tab2} and \ref{tab3}. The emission across all wavelengths is optically thick (i.e., the size of the atmosphere continuously decreases as frequency increases) and thermal, so that the brightness temperature is just the local gas temperature  where the optical depth is near unity, averaged across the stellar disk. In Figure \ref{fig3} we plot these gas temperatures against projected distance from the optical photosphere, where the diameter of the optical photosphere is 37.38 mas \citep{ohnaka_2013}. The highest ALMA frequency we have at 405 GHz probes the inner region of the extended atmosphere of Antares at 1.35 R$_{\star}$ (i.e., 0.35 R$_{\star}$ above the photosphere) and provides a measure of the gas temperature of 2682 $\pm$ 400 K, where the uncertainties in the gas temperature are dominated by the conservative uncertainties in the absolute flux calibration. The lower ALMA frequencies and highest VLA frequencies show that the gas temperature then progressively rises to a maximum value of 3777 $\pm$ 567 K at 2.5 R$_{\star}$. This continuous rise in the gas temperature is the first direct evidence for the existence of a chromospheric temperature rise in the atmosphere of an RSG. The single ALMA measurement by \cite{ogorman_2017} for Betelgeuse indicated this temperature rise, but the sampling was too sparse and error bars too large to conclusively show it. We note that in classical 1D chromospheric models based on optical and UV emission line studies \citep[e.g.,][]{basri_1981}, the temperature at the bottom of the chromosphere is often defined as T$_{\rm{min}} \sim 0.75T_{\rm{eff}}$. For Antares, this would give T$_{\rm{min}} \sim 2745\,$K, which is very similar to our ALMA band 8 temperature value. Our ALMA and VLA dataset may therefore have sampled the entire chromosphere of Antares. The temperature of the Antares atmosphere progressively falls to lower values beyond 2.5 R$_{\star}$ until it reaches a value of 1647 $\pm$ 110 K at 11.5 R$_{\star}$, in a similar manner as was reported by \cite{lim_1998} and \cite{ogorman_2015} for Betelgeuse. 

It appears that the different spectral index values discussed in Section \ref{sec3.2} can be equated to the temperature profile plotted in Figure \ref{fig3} by a comparison of values on a per frequency basis.  The spectral index value of 1.42 derived from the ALMA and highest VLA frequencies now appears to originate from the chromosphere. The spectral index value of 0.80 derived at the lowest VLA frequencies appears to emanate from the wind where it is approaching the terminal velocity at the current epoch, as measured by the wide strong K\,I absorption $\sim$20\,km\,s$^{-1}$ \citep{sanner_1976, pugh_2013}.  Finally, the spectral index turnover between 10 and 20 GHz could then probe the wind acceleration region. 

\section{Discussion}
\subsection{Reconciliation of the chromosphere and MOLsphere}

A CO molecular envelope (known as a MOLsphere) lying above the classical photosphere has been imaged for Antares by means of near-infrared spectro-interferometric observations in the CO first-overtone lines near 2.3 $\mu$m \citep{ohnaka_2013}. The modeling of Ohnaka and colleagues suggests that the CO MOLsphere lies between 1.2-1.4 R$_{\star}$. This extent is similar to that observed in Betelgeuse in the 2.3 $\mu$m CO lines \citep{ohnaka_2011, tsuji_2006} and in water vapor \citep{tsuji_2000, perrin_2004, ohnaka_2004, tsuji_2006, perrin_2007, Montarges_2014}. The radius of the MOLsphere corresponds to the region between the upper photosphere and the radio temperature peak, reminiscent of the CO material observed at low chromospheric heights on the Sun and thought to exist in cool evolved stars \cite[e.g.,][]{wiedemann_1994}. 

The temperature of the MOLsphere of both Antares and Betelgeuse is estimated to be $\sim$2000~K at 1.2--1.4\,$R_{\star}$ based on the modeling of spectroscopic and interferometric data mentioned above. These temperatures are noticeably lower than the gas temperature derived at the same radii from the ALMA observations. Based on the modeling of the optically thin [Fe II] line at 17.94 $\mu$m for Betelgeuse, \cite{harper_2009} also derived a gas temperature $<$2500\,K at 0".035 = 1.7\,$R_{\star}$. Furthermore, the recent near-infrared polarimetric aperture-masking observations of Betelgeuse by \cite{houbois_2019} reveal dust formation at close to 1.5\,R$_{\star}$. This means that the temperature at 1.5\,R$_{\star}$ should be lower than the often adopted dust condensation temperature of ~1500\,K, while the gas temperature measured with ALMA at the same radius is as high as 3000\,K. 

The radial thermal structure of the Antares atmosphere derived from the ALMA and VLA data represents the mean of any inhomogeneous components across the stellar disk within each radio beam. Therefore it might be argued that the temperature derived from the ALMA and VLA data is the average of the hot (i.e., $\sim$7000\,K \citealt{basri_1981}) component responsible for the UV emission lines and the cooler MOLsphere ($\sim$2000~K) and the dust-forming gas ($<$ 1500~K). However, we estimate the optical depth of the MOLsphere to be much lower than unity, mainly due to H$^{-}$ opacity, as shown in Appendix \ref{app1}. This means that the MOLsphere (and also the dust-forming cool gas) is invisible across all ALMA bands, and it does not contribute to the measured temperature. Therefore the \textit{\textup{lukewarm}} chromosphere, whose temperature reaches 3000~K at $\sim$1.5 R$_{\star}$, does not represent the average of the hot and cool components. It is unlikely that the distribution of the hot $(T_e > 7000$\,K) chromospheric components occupies a large area fraction of the stellar disk because there are only minor intensity features in the uniform disk-subtracted residual images (see Appendix \ref{app1}). Moreover, the hottest gas detected in the very high resolution (i.e., 14\,mas) image of Betelgeuse by \cite{ogorman_2017} was $\sim 3800\,$K, which was still well below typical hot chromospheric values and only occupied a few percent of the projected disk area. The area filling factor of the hot chromospheric gas is therefore probably smaller than or indeed far smaller than $10^{-2}$.

It is possible that the MOLsphere and the dust-forming cool gas coexist with the lukewarm chromosphere. This latter component should then be dominant and uniformly distributed across the beam because otherwise, the angular sizes at the ALMA bands would be noticeably smaller than those observed. However, if CO molecules are present at 3000~K and are not photodissociated by the embedded hot FUV-emitting plasma component \citep[e.g.,][]{visser_2009}, the dominant lukewarm chromosphere extending to 1.3--1.5~R$_{\star}$ would lead to strong emission of the 2.3~$\mu$m CO lines off the limb of the star. This emission would fill in the absorption expected over the photospheric disk, making the CO lines appear significantly weaker than those observed. Furthermore, the images of Antares recently obtained from the 2.3~$\mu$m CO lines with an angular resolution of 5.4~mas, which is seven times finer than the stellar angular diameter, show that the MOLsphere extends out to $\sim$1.7 R$_{\star}$ and does not appear to be very patchy, although there are some inhomogeneities \citep{ohnaka_2017}. The interferometric and spectroscopic observations of the 2.3~$\mu$m CO lines therefore suggest that the lukewarm chromosphere cannot be the dominant component if CO has a significant abundance at 3000\,K. This appears to contradict the ALMA measurements.

A picture to reconcile these results is that the lukewarm chromosphere and the cool component (MOLsphere and dust-forming cool gas) exist in separate structures whose spatial scales are smaller than the angular resolution of the 2.3~$\mu$m CO line images mentioned above, perhaps analogously to the formation of H$\alpha$ in cool giants \citep{eaton_1995}. In this case, the fine inhomogeneous structures do not affect the overall size of the outer atmosphere seen at the ALMA bands and in the 2.3~$\mu$m CO lines. Such inhomogeneous structures is a key for better models of the outer atmosphere of RSGs. For example, while the aforementioned infrared interferometric studies reveal the presence of the MOLsphere, the current MOLsphere models cannot properly explain the 12~$\mu$m H$_2$O lines observed in absorption, as pointed out by \cite{ryde_2006}: the H$_2$O lines are predicted to be too weak in absorption or even in emission. The coexistence of the lukewarm chromosphere, MOLsphere, and the dust-forming cool gas may reconcile this problem. 

\subsection{NLTE modeling of the FUV radiation field}\label{sec_nonlte}
\begin{figure}[t]
\centering 
\includegraphics[angle=90, scale=0.41]{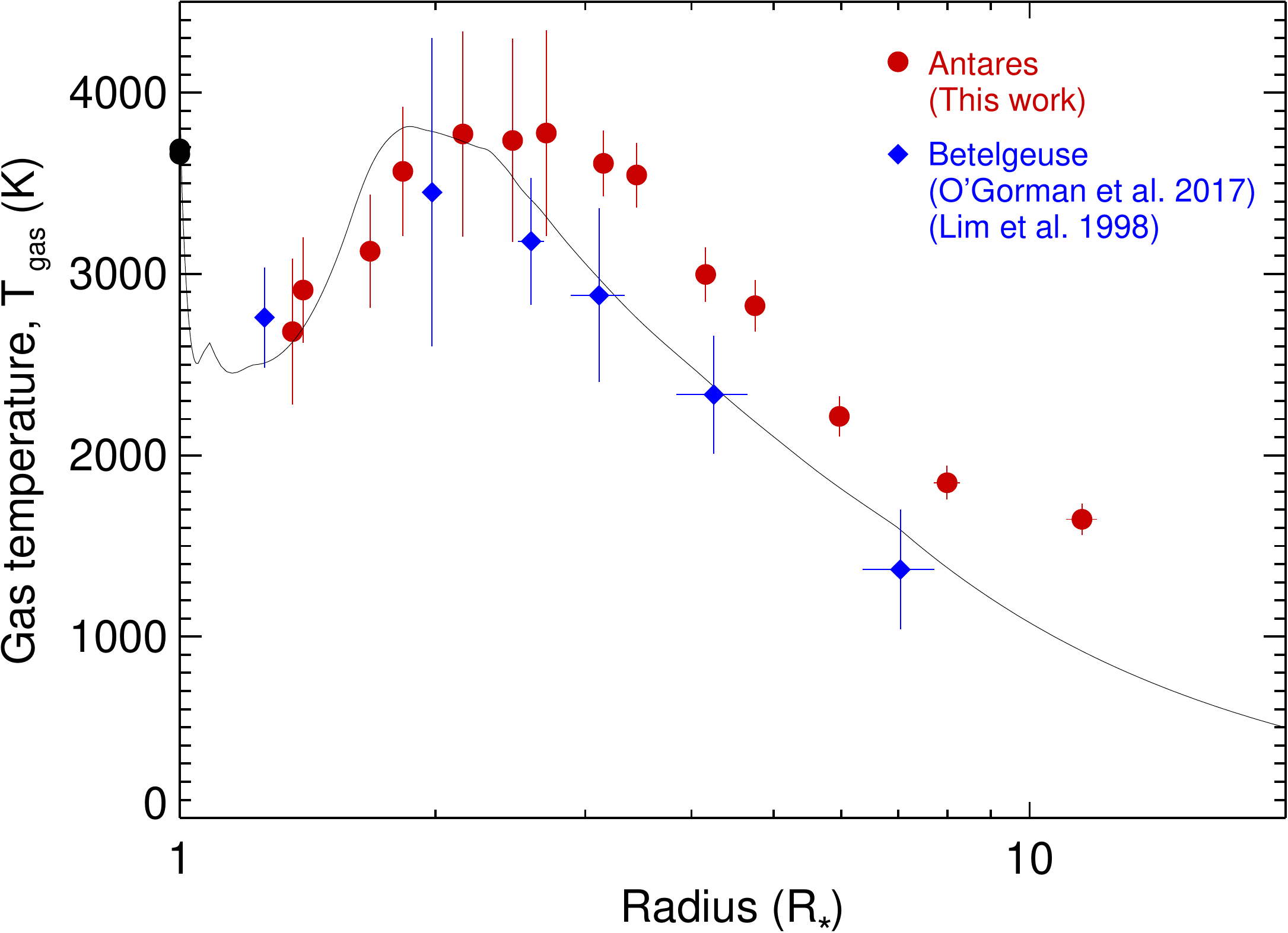}
\caption[]{Comparison of the temperature structure of the extended atmospheres of the early-M supergiants Antares and Betelgeuse. The filled red circles are the measurements summarized in this work for Antares, and the filled blue diamonds are the measurements for Betelgeuse \citep{lim_1998, ogorman_2017}. The error bars in the gas temperature include the uncertainty in absolute flux density scale. The overlapping filled black circles at R$_{\star} = 1$ represent the photospheric effective temperatures of Antares and Betelgeuse. The solid black line is the temperature profile of the semiempirical model for the extended atmosphere of Betelgeuse discussed in Section \ref{sec_nonlte} and Appendix \ref{app1}.  }
\label{fig3a}
\end{figure}

The FUV spectrum of RSGs is an independent probe of the hottest material in their atmospheres. It is therefore interesting to investigate whether the chromospheric properties now detected at radio wavelengths are sufficient to explain what is observed in the FUV. An understanding of the FUV spectrum of RSGs is indeed essential for determining the electron, ion, and molecular abundances, which in turn control the dust formation. While  {\it Copernicus} spectra of the chromospheric near-UV Mg\,II h and k emission lines \citep{bernat_1976} reveal that the surfaces fluxes of Antares and Betelgeuse are very similar when modern angular diameters are adopted, the presence of the hot B-type companion of Antares in IUE spectra and the absence of an HST spectrum has meant that the best available M supergiant FUV spectrum is that of Betelgeuse. The following analysis is therefore focused on Betelgeuse alone. However, the now confirmed similarities of the temperature profiles of their extended atmospheres, demonstrated in Figure \ref{fig3a}, makes the discussion appropriate for both stars, and probably all early-M RSGs.

The FUV radiation field of Betelgeuse has been observed at high signal-to-noise ratio (S/N) with the Goddard High Resolution Spectrograph (GHRS) \citep{carpenter_1994} and the Space Telescope Imaging Spectrograph (STIS) of the HST \citep{carpenter_2018}, and it is interesting to compare these observations with the FUV continuum generated by empirically derived VLA and ALMA radio temperature profiles. To do so, we performed time-independent nonlocal thermal equilibrium (NLTE) spherical radiative transfer computations for the formation of the Si~I continua to the ground and first excited terms, which are expected to dominate the bound-free opacity between $1350 \textless \lambda {\rm (\AA)} \textless 1680$. Radio continuum bremsstrahlung opacity is proportional to the electron density \citep{rybicki_1986, dalgarno_1966} from singly ionized abundant metals of low first ionization potential (e.g., Si, Fe, and Mg) whose ground-state photoionization edges $< 1625$\AA. We used a revision of the \cite{harper_2001} semiempirical thermodynamic model for Betelgeuse that is based on the ALMA and VLA temperature distribution shown in Figure 2 of \cite{ogorman_2017}. The details of this new model and the our FUV simulations will be presented elsewhere and are summarized in Appendix \ref{app2}. 

\begin{figure}[t!]
\centering 
\includegraphics[angle=0, scale=0.41]{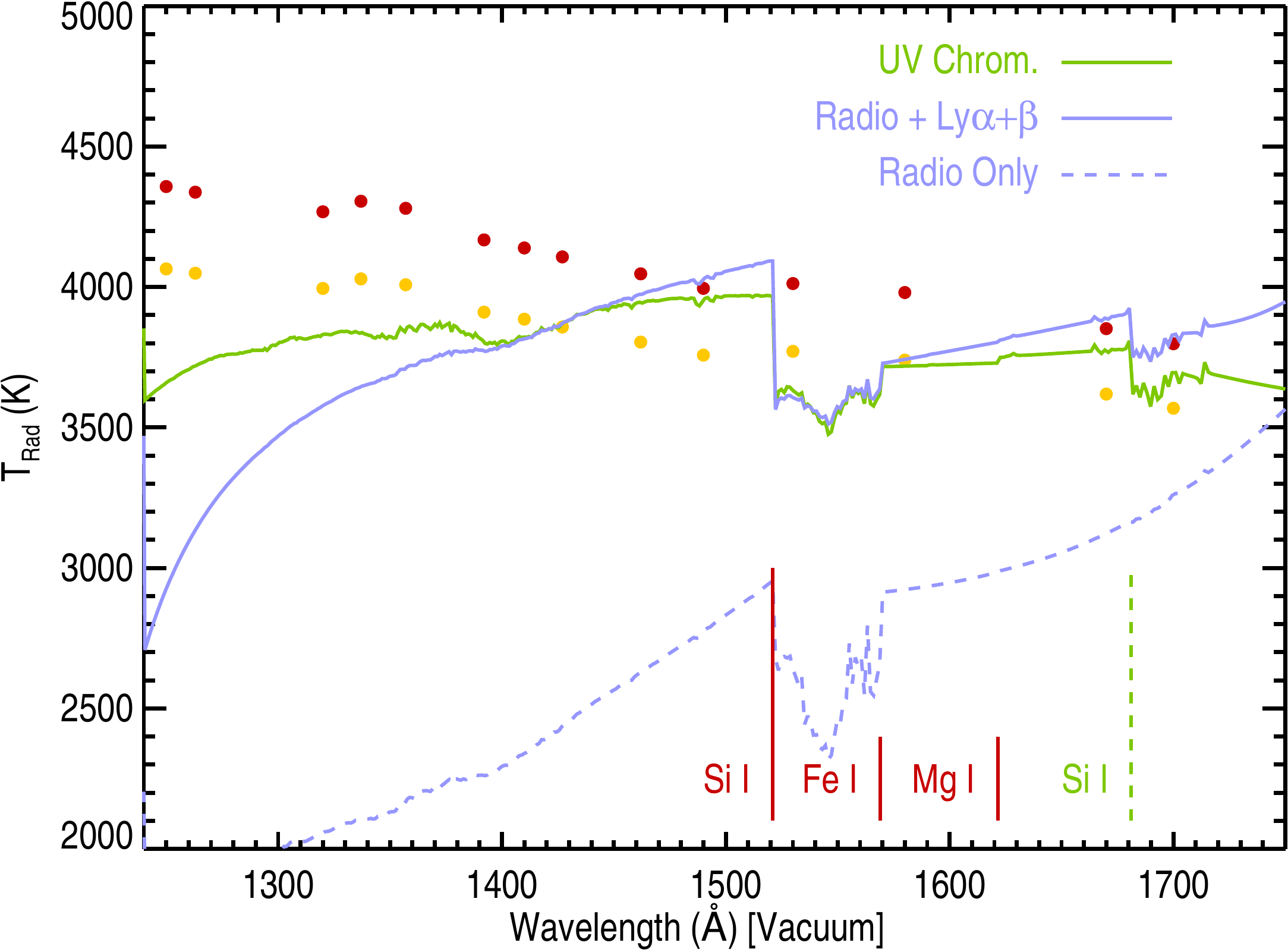}
\caption[]{Observed and simulated radiation temperatures for the FUV spectrum of Betelgeuse. The observed continuum values are shown as solid circles: yellow for no ISM reddening, and red for an extinction of $A_V=0.62$ (see text for details). The blue lines are simulations based on an empirically derived temperature profile of the extended atmosphere of Betelgeuse. The dashed line is the line predicted in the absence of any additional sources of excitation. Si is partially ionized, and there is sufficient opacity for the ground-state edge to be optically thick in the cooler outer layers, leading to very low fluxes $\lambda < 1680$\AA. Adding photoionization from H~I Ly$\alpha$ and Ly$\beta$ increases the ionization to the point where the optical depth unity surface is near the thermal peak at $2R_\star$. The solid green line is the line from the compact hot chromospheric model of \cite{basri_1981}. The hydrogen ionization from the hot chromospheric plasma leads to additional emission at shorter wavelengths. The solid vertical red lines indicate the location of the ground-state photoionization edges, and the dashed green line is the photoionization edge form the first excited Si~I term. The Si~I edges are intrinsically stronger than those of Fe~I and Mg~I.}
\label{fig4}
\end{figure}

Figure \ref{fig4} shows the Betelgeuse STIS observations as radiation temperatures, assuming an angular diameter of 44\,mas, for the cases of (a) no interstellar medium (ISM) reddening: yellow filled circles ($A_V=0.00$), and (b) a reddening of $A_V=0.62$ with $R_V=3.1$ (e.g., \citealt{levesque_2005}) using the FUV extinction relation of \cite{cardelli_1989}: red filled circles. The continuum flux points are presented as radiation temperatures, $T_{Rad}$, which are defined as the temperature required by an isotropic blackbody to produce the observed surface flux density. This temperature is defined as
\begin{equation}\label{eq1}
  T_{Rad} =   {hc\over{k\lambda}}\left\{
  \ln\left[\frac{hc^2\pi \phi_\star^2}{2F_{\lambda\oplus}\lambda ^5 } +1\right]\right\}^{-1}
,\end{equation}
where $F_{\lambda\oplus}$ is the observed flux, $\phi_\star$ is the photospheric angular diameter, and the other symbols have their usual meanings. The continuum flux values were chosen to avoid the narrow circumstellar CO fourth-positive absorption bands and chromospheric emission lines. Figure \ref{fig4} shows that starting from 1700\AA, the observed $T_{Rad}$ slowly increases toward shorter wavelengths \citep{carpenter_2018}, as is observed for the Sun (\citealt{avrett_2008}, Fig.~1), albeit at lower values.

The revised thermodynamic model of \cite{harper_2001} predicts the $T_{Rad}$ distribution shown as the dashed blue line. Given that $T_{rad}$ is logarithmically dependent on the observed flux density (see Equation \ref{eq1}), the discrepancy between the observation and predictions is enormous. In this model, silicon is predominantly partially ionized, and the continuum source function is only weakly coupled to the thermal structure. However, the ground-state continuum is optically thick in the cool outer reaches. Embedded in the cool extended atmosphere that was first detected by \cite{lim_1998} lies hot chromospheric plasma that generates strong collisionally excited UV emission lines (e.g., \citealt{harper_2006}). We therefore included photoionization by H~I Ly$\alpha$ and Ly$\beta$, assuming that the intrinsic integrated H~I Ly$\alpha$ and Mg~II h \& k fluxes are the same (based on \citealt{wood_2005}) and that the Ly$\beta$ flux is assumed to be 1/250 that of  Ly$\alpha$. The solid blue line shows a much closer agreement at longer wavelengths. In this model, Si~I is predominately singly ionized, and the source function is much larger and even less coupled to the thermal structure. The ground-state opacity is reduced, leading to enhanced FUV emission. However, shortward of 1400\AA,{ } the FUV radiation field remains very strongly underestimated, which is  partly the result of excess opacity in the wing of H Ly$\alpha$.

As a comparison, we computed the FUV continuum using the hot compact chromospheric model of \cite{basri_1981}. This model was partially based on UV chromospheric emission lines measured with the International Ultraviolet Explorer, {\it IUE}. The resulting spectrum is shown in Figure \ref{fig4}. This model, which includes no cool extended plasma, remarkably provides a crude match to the observed spectrum, given the uncertainties in reddening from the ISM and circumstellar dust (see, e.g., \citealt{kervela_2011}). This illustrates the perils of the monochromatic approach to studying the atmospheres of cool stars, especially noncoronal red supergiants, because the presence of cool extended plasma must be accounted for. In terms of computing the radio opacity in the extended atmosphere, adopting the radio thermal structure and including an additional chromospheric photoionization component appears to yield reasonable estimates of the singly ionized abundances. Proceeding on simultaneously matching the radio continuum and the FUV continuum will require multidimensional atmospheric models that explicitly include the hot localized plasma, which will increase the flux shortward of 1400\AA. The UV-emitting plasma might be trapped in magnetic fields that are heated and become ionized, or it may originate in unresolved (small-scale) shock fronts that temporarily heat and ionize the gas.

\subsection{Constraints on the maximum chromospheric temperature}
An examination of the STIS E230M spectrum of Betelgeuse in the HST ASTRAL spectroscopic library [P.I. T. Ayres https://archive.stsci.edu/prepds/astral/\#coolstars, \cite{ayres_2010}] reveals the optically thin Si III] 1892$\AA$ emission line. An emission measure analysis (see, e.g., \cite{brown_1984}) with the nearby Si II 1808+1816$\AA$ multiplet, and also Al II] 2669$\AA$, Mg II h 2802$\AA,$ and the nondetection of C III] 1908$\AA$, indicates that chromospheric plasma must be heated to above 12,500\,K to account for the Si III] emission. The strongest constraint is from the ratio of the Si II and Si III] fluxes because they have the same elemental abundance and very similar ISM and circumstellar reddening. A weak constraint from the C III] nondetection is that the maximum temperature is $< 60,000$K.

\subsection{Heating of the chromosphere}
The VLA and ALMA data demonstrate their synergy by enabling a complete frequency sweep for temperature tomography. The distribution shown in Figure \ref{fig3} is more complete and has a higher S/N than was achieved for Betelgeuse \citep{ogorman_2017}. The similarity of the two thermal structures indicates that this is a common phenomenon in early-M RSGs and not just a quirk of individual stars.

The temperature peak near 2.5\,R$_\star$ is similar in extent to that in Betelgeuse, and it is interesting to compare this extent  with that derived from the most detailed 1D semiempirical model of an RSG, that is, $\alpha$~Tau (K5 III) \citep{mcmurry_1999}. This model was based on {\it Hubble} FUV and near-UV spectra and included turbulent and gas pressure support. It rises from a temperature minimum to transition region temperatures at the outer boundary at a fractional extent of $\Delta R/R_\star \simeq 20\%$.

In 1D semiempirical chromospheric models of nonpulsating cool evolved stars, one of the characteristic properties is that the electron density, $n_e$, remains reasonably constant with height (to within a factor of $\pm 3$) while the total hydrogen density ($n_H$ = HI + HII) falls by four to five orders of magnitude. In order for the hydrogen ionization fraction to increase outward, the gas temperature must also increase, indicating that the heating rate per gram increases outward. This implies that the chromospheric heating per gram falls off more slowly than the hydrogen density \citep[cf.][]{ayres_1979}. At the temperature minimum, the source of electrons is photoionization of the lowest first-ionization potential metals:
K, Ca, Na, Al, Si, Fe, and Mg, and $x_{met} = n_e/n_H \simeq 10^{-4}$, and at higher temperatures, C and H become ionized.

As the hydrogen density falls, the heating exceeds the cooling and the gas temperature increases, and hydrogen begins to contribute electrons by photoionization
by the Balmer continuum, following excitation of the $n=2$ level \citep{hartmann_1984}. When the hydrogen
density has declined by $1/x_{met}$ , it becomes dominantly ionized at $T_{gas}\sim 10^{4} \,$K, and then the source of available electrons declines rapidly as the hydrogen density continues to fall. When we apply this concept to the embedded hot chromospheric component of M supergiants, we can establish the radius where this occurs, $R_{top}$, by using the density stratification in a thermal and turbulent supported atmosphere (e.g, see \citealt{eaton_1993}),
\begin{equation}
 \ln\left\{ x_{met} \right\} =
 \ln\left\{ {n_H (R_{top}) \over{ n_H (R_\star)}} \right\}
 = {R_\star^2\over{H_\star}} \left( {1\over{R_{top}}} - {1\over{R_\star}}\right)
 \label{eq:rmax}
,\end{equation}
where the density scale height at the stellar surface, $H_\star$ (assumed constant), is given by
\begin{equation}
H_\star = \left({kT_{gas}\over{\mu m_H}} + {1\over{2}}\rm{v}_{turb}^2\right)/g_\star.
\end{equation}
$\rm{v}_{turb}$ is the most probable turbulent velocity, assumed
be isotropic and to have spatial scales smaller than $H_\star$.

We adopted $M_\star=15$\,M$_{\odot}$
and $R_\star=680$\,R$_\odot$ \citep{ohnaka_2013}, with a mid-chromospheric
$T_{gas}=8000$\,K and  $\rm{v}_{turb}=19\>{\rm km\>s}^{-1}$ measured from Betelgeuse
(Harper and Brown 2006). Evaluating Eq.~(\ref{eq:rmax}), we find
$R_{top}=2.0\,R_\star$. There is some indication that the turbulence in the extended atmosphere of Antares is greater than in Betelgeuse (\citealt{harper_2009}, Table 3), which
would lead to a greater extension in Antares, as observed. When we make the same evaluation for $\alpha$~Tau, we find
$R_{top}=1.2\,R_\star$. Given the uncertainty in surface gravity (30\%), and
because we took a constant $T_{gas}$ and $\rm{v}_{turb}$ in the density scale height, these results are consistent with the assumption that the heating rate per gram increases outward within a
predominantly turbulent supported chromosphere.

\subsection{RSGs versus AGB stars at radio wavelengths}
Some progress has recently been made into understanding the physical properties of asymptotic giant branch (AGB) stars at radio wavelengths, and  it is therefore valuable to compare their properties at these wavelengths to those of the RSGs. \cite{reid_1997} detected six AGB stars between 6 and 22\,GHz and found that the emission was characteristic of the Rayleigh-Jeans tail of a thermal blackbody source (i.e., $S_{\nu} \propto \nu ^2$), with their flux densities exceeding those expected from their optical photospheres by roughly a factor of 2. The authors also successfully spatially resolved one of their targets (i.e., W Hya) at 22\,GHz and found that its diameter was twice that of the photospheric angular diameter, although its brightness temperature was $\sim 1000\,$K below the photospheric effective temperature (T$_\textrm{eff}$ $\sim$ 2500\,K). They coined the term \textit{\textup{radio photosphere}} to describe the region of the atmosphere at $\sim 2\,$R$_{\star}$ from which the radio emission emanates, with the opacity being generated from electrons from metals with low-ionization potentials. These properties for the radio photospheres of AGB stars have now been confirmed by multiple subsequent studies \citep{reid_2007, menten_2012, matthews_2015, planesas_2016, matthews_2018, vlemmings_2019}.

 Although the term radio photosphere was initially only used to describe the radio atmospheres of AGB stars, it is now sometimes used in the literature to describe the radio atmospheres of RSGs \citep[e.g.,][]{carilli_2018}. Our results here have shown that the radio properties of the RSGs are very different to those of AGB stars. For example, unlike in AGB stars, the radio spectral index for the RSGs is much more shallow than the Rayleigh-Jeans tail of a thermal blackbody source between at least 3 and 405\,GHz. Moreover, we have shown that radio emission from an RSG between roughly 22 and 400\,GHz traces optically thick chromospheric emission, while no evidence for a chromospheric temperature rise is found by \cite{matthews_2015} over a similar frequency range for the AGB star Mira A. Finally, at frequencies lower than 22\,GHz for the RSGs, the relatively low spectral index values ($\sim$\,0.8) along with the declining gas temperature indicate that this radio emission stems from an expanding wind. However, for AGB stars at these frequencies, the spectral index is still $\sim 2,$ which would indicate that the presumably less ionized wind is still too optically thin to be detectable. We conclude that the inclusion of the term \textit{\textup{radio photosphere}} is unnecessary when the chromosphere and wind of an RSG at radio wavelengths is described and might indeed be confused with either the classical optical photosphere of an RSG or with the different properties of the AGB radio photospheres.
\section{Conclusions and future prospects}
We have performed an extensive spatially resolved study of the thermal free-free continuum emission from the RSG Antares at centimeter to submillimeter wavelengths. To the best of our knowledge, this is the first time that a stellar disk other than the solar disk has been spatially resolved longward of 6\,cm. Our fine frequency sampling between 405\,GHz and 3\,GHz along with the high dynamic range of our data has allowed us to locate the chromospheric temperature rise and onset of the wind in the atmosphere of an RSG. Our dataset demonstrates the power of spatially resolved mutifrequency radio continuum observations for understanding the nature of evolved stellar atmospheres. 

Our study also highlights the importance of including observations at different parts of the electromagnetic spectrum to acquire as much of an understanding of the nature of RSG atmospheres as possible. To reconcile previously reported hot (i.e., $\sim$7,000-9,000\,K) FUV emission studies of RSGs with our radio data requires that this hot FUV-emitting plasma must have a very small filling factor throughout the chromosphere in order for us to measure much lower temperatures at radio wavelengths. A similar conclusion was drawn by \cite{lim_1998} for the more extended atmosphere of Betelgeuse beyond $\sim$2.5\,R$_{\star}$. We therefore picture a lukewarm chromosphere that peaks at $3800\,$K at $\sim$2.5R$_{\star}$ and is speckled with hotter FUV-emitting plasma throughout. The cool (i.e., $<$2000\,K) MOLspheres around RSGs that have been extensively studied at near-IR wavelengths are optically thin across all of the wavelengths reported in this study and thus do not contribute to our derived temperature profile. This means that the cool MOLsphere must also coexist with our lukewarm chromosphere on spatial scales below the spatial resolution of current telescopes.

Bright chromospheric substructures have previously been imaged and resolved at 338\,GHz with ALMA for Betelgeuse \citep{ogorman_2017} and indicate localized regions of enhanced heating. We compared the similarities of the temperature profiles in the atmospheres of Betelgeuse and Antares, and our results show that the substructures imaged for Betelgeuse are present in its lower chromosphere. It would be intriguing to carry out further high-resolution imaging with ALMA of either Antares and/or Betelgesue over a number of epochs  to monitor the prevalence and evolution of these substructures to determine the timescales upon which they evolve. A comparison of these timescales to known timescales, such as those of rotation or convection, might help to explain their origin. A repetition of these high-resolution ALMA observations at multiple wavelengths might then tell us to which extent these regions of enhanced heating reach out into the atmosphere. The 15\% flattening of the atmosphere that we observed here for Antares at all four ALMA wavelengths might indicate that they permeate much of the chromosphere. Future radio interferometers such as the next-generation Very Large Array and the Square Kilometre Array will probe the very top of the chromosphere along with the wind-launching region at a fidelity and resolution similar to those that have been achieved for the lower chromosphere of Betelgeuse with ALMA.

\begin{acknowledgements}
The data presented in this paper were obtained with the Karl G. Very Large Array (VLA) which is an instrument of the National Radio Astronomy Observatory (NRAO). The National Radio Astronomy Observatory is a facility of the National Science Foundation operated under cooperative agreement by Associated Universities, Inc. This paper makes use of the following ALMA data: ADS/JAO.ALMA\#2016.1.00234.S. ALMA is a partnership of ESO (representing its member states), NSF (USA) and NINS (Japan), together with NRC (Canada), MOST and ASIAA (Taiwan), and KASI (Republic of Korea), in cooperation with the Republic of Chile. The Joint ALMA Observatory is operated by ESO, AUI/NRAO and NAOJ. GMH received support from Program number HST-AR-14566 which was provided by NASA through a grant from the Space Telescope Science Institute, which is operated by the Association of Universities for Research in Astronomy, Incorporated, under NASA contract NAS5-26. KO acknowledges the support of the Comisión Nacional de Investigación Científica y Tecnológica (CONICYT) through the FONDECYT Regular grant 1180066. A. F.-J.  would like to acknowledge support from the European Research Council grant number 743029.
\end{acknowledgements}



\bibliographystyle{aa}
\bibliography{references}

\begin{thebibliography}{88}
\expandafter\ifx\csname natexlab\endcsname\relax\def\natexlab#1{#1}\fi

\bibitem[{{Asplund} {et~al.}(2009){Asplund}, {Grevesse}, {Sauval}, \&
  {Scott}}]{asplund_2009}
{Asplund}, M., {Grevesse}, N., {Sauval}, A.~J., \& {Scott}, P. 2009, \araa, 47,
  481

\bibitem[{{Avrett} \& {Loeser}(2008)}]{avrett_2008}
{Avrett}, E.~H. \& {Loeser}, R. 2008, \apjs, 175, 229

\bibitem[{{Ayres}(1979)}]{ayres_1979}
{Ayres}, T.~R. 1979, \apj, 228, 509

\bibitem[{{Ayres}(2010)}]{ayres_2010}
{Ayres}, T.~R. 2010, \apjs, 187, 149

\bibitem[{{Basri} {et~al.}(1981){Basri}, {Linsky}, \& {Eriksson}}]{basri_1981}
{Basri}, G.~S., {Linsky}, J.~L., \& {Eriksson}, K. 1981, \apj, 251, 162

\bibitem[{{Bernacca} \& {Perinotto}(1973)}]{bernacca_1973}
{Bernacca}, P.~L. \& {Perinotto}, M. 1973, {A catalogue of stellar rotational
  velocities. Vol.1: Main sequence single stars. Vol.2: Main sequence
  spectroscopic binaries; Vol.4: Evolved stars}

\bibitem[{{Bernat} \& {Lambert}(1976)}]{bernat_1976}
{Bernat}, A.~P. \& {Lambert}, D.~L. 1976, \apj, 204, 830

\bibitem[{{Boesgaard} \& {Boesgaard}(1976)}]{boesgaard_1976}
{Boesgaard}, A.~M. \& {Boesgaard}, H. 1976, \apj, 205, 448

\bibitem[{{Braun} {et~al.}(2012){Braun}, {Baade}, {Reimers}, \&
  {Hagen}}]{braun_2012}
{Braun}, K., {Baade}, R., {Reimers}, D., \& {Hagen}, H.~J. 2012, \aap, 546, A3

\bibitem[{{Brown} \& {Harper}(2004)}]{brown_2004}
{Brown}, A. \& {Harper}, G.~M. 2004, in IAU Symposium, Vol. 219, Stars as Suns
  : Activity, Evolution and Planets, ed. A.~K. {Dupree} \& A.~O. {Benz}, 646

\bibitem[{{Brown} {et~al.}(1984){Brown}, {Jordan}, {Stencel}, {Linsky}, \&
  {Ayres}}]{brown_1984}
{Brown}, A., {Jordan}, C., {Stencel}, R.~E., {Linsky}, J.~L., \& {Ayres}, T.~R.
  1984, \apj, 283, 731

\bibitem[{{Cardelli} {et~al.}(1989){Cardelli}, {Clayton}, \&
  {Mathis}}]{cardelli_1989}
{Cardelli}, J.~A., {Clayton}, G.~C., \& {Mathis}, J.~S. 1989, \apj, 345, 245

\bibitem[{{Carilli} {et~al.}(2018){Carilli}, {Butler}, {Golap}, {Carilli}, \&
  {White}}]{carilli_2018}
{Carilli}, C.~L., {Butler}, B., {Golap}, K., {Carilli}, M.~T., \& {White},
  S.~M. 2018, in Astronomical Society of the Pacific Conference Series, Vol.
  517, Science with a Next Generation Very Large Array, ed. E.~{Murphy}, 369

\bibitem[{{Carpenter} {et~al.}(2018){Carpenter}, {Nielsen}, {Kober}, {Ayres},
  {Wahlgren}, \& {Rau}}]{carpenter_2018}
{Carpenter}, K.~G., {Nielsen}, K.~E., {Kober}, G.~V., {et~al.} 2018, \apj, 869,
  157

\bibitem[{{Carpenter} {et~al.}(1994){Carpenter}, {Robinson}, {Wahlgren},
  {Linsky}, \& {Brown}}]{carpenter_1994}
{Carpenter}, K.~G., {Robinson}, R.~D., {Wahlgren}, G.~M., {Linsky}, J.~L., \&
  {Brown}, A. 1994, \apj, 428, 329

\bibitem[{{Carr} {et~al.}(2000){Carr}, {Sellgren}, \&
  {Balachandran}}]{carr_2000}
{Carr}, J.~S., {Sellgren}, K., \& {Balachandran}, S.~C. 2000, \apj, 530, 307

\bibitem[{{Cunto} {et~al.}(1993){Cunto}, {Mendoza}, {Ochsenbein}, \&
  {Zeippen}}]{cunto_1993}
{Cunto}, W., {Mendoza}, C., {Ochsenbein}, F., \& {Zeippen}, C.~J. 1993, A\&A,
  275, L5

\bibitem[{{Dalgarno} \& {Lane}(1966)}]{dalgarno_1966}
{Dalgarno}, A. \& {Lane}, N.~F. 1966, \apj, 145, 623

\bibitem[{{De Koter} {et~al.}(1988){De Koter}, {de Jager}, \&
  {Nieuwenhuijzen}}]{de_koter_1988}
{De Koter}, A., {de Jager}, C., \& {Nieuwenhuijzen}, H. 1988, \aap, 200, 146

\bibitem[{{Eaton}(1993)}]{eaton_1993}
{Eaton}, J.~A. 1993, \apj, 404, 305

\bibitem[{{Eaton}(1995)}]{eaton_1995}
{Eaton}, J.~A. 1995, \aj, 109, 1797

\bibitem[{{Fomalont} {et~al.}(2014){Fomalont}, {van Kempen}, {Kneissl},
  {Marcelino}, {Barkats}, {Corder}, {Cortes}, {Hills}, {Lucas}, {Manning}, \&
  {Peck}}]{fonalont_2014}
{Fomalont}, E., {van Kempen}, T., {Kneissl}, R., {et~al.} 2014, The Messenger,
  155, 19

\bibitem[{{Fontenla} {et~al.}(2011){Fontenla}, {Harder}, {Livingston}, {Snow},
  \& {Woods}}]{fontenla_2011}
{Fontenla}, J.~M., {Harder}, J., {Livingston}, W., {Snow}, M., \& {Woods}, T.
  2011, Journal of Geophysical Research (Atmospheres), 116, D20108

\bibitem[{{Gavrila}(1967)}]{gavrila_1967}
{Gavrila}, M. 1967, Physical Review, 163, 147

\bibitem[{{Gilliland} \& {Dupree}(1996)}]{gilliland_1996}
{Gilliland}, R.~L. \& {Dupree}, A.~K. 1996, \apjl, 463, L29

\bibitem[{{Gustafsson} {et~al.}(2008){Gustafsson}, {Edvardsson}, {Eriksson},
  {J{\o}rgensen}, {Nordlund}, \& {Plez}}]{gustafsson_2008}
{Gustafsson}, B., {Edvardsson}, B., {Eriksson}, K., {et~al.} 2008, \aap, 486,
  951

\bibitem[{{Harper}(1994)}]{harper_1994}
{Harper}, G.~M. 1994, \mnras, 268, 894

\bibitem[{{Harper} \& {Brown}(2006)}]{harper_2006}
{Harper}, G.~M. \& {Brown}, A. 2006, \apj, 646, 1179

\bibitem[{{Harper} {et~al.}(2017){Harper}, {Brown}, {Guinan}, {O'Gorman},
  {Richards}, {Kervella}, \& {Decin}}]{harper_2017}
{Harper}, G.~M., {Brown}, A., {Guinan}, E.~F., {et~al.} 2017, \aj, 154, 11

\bibitem[{{Harper} {et~al.}(2001){Harper}, {Brown}, \& {Lim}}]{harper_2001}
{Harper}, G.~M., {Brown}, A., \& {Lim}, J. 2001, \apj, 551, 1073

\bibitem[{{Harper} {et~al.}(2009){Harper}, {Richter}, {Ryde}, {Brown}, {Brown},
  {Greathouse}, \& {Strong}}]{harper_2009}
{Harper}, G.~M., {Richter}, M.~J., {Ryde}, N., {et~al.} 2009, \apj, 701, 1464

\bibitem[{{Harris} \& {Lambert}(1984)}]{harris_1984}
{Harris}, M.~J. \& {Lambert}, D.~L. 1984, \apj, 281, 739

\bibitem[{{Hartmann} \& {Avrett}(1984)}]{hartmann_1984}
{Hartmann}, L. \& {Avrett}, E.~H. 1984, \apj, 284, 238

\bibitem[{{Haubois} {et~al.}(2019){Haubois}, {Norris}, {Tuthill}, {Pinte},
  {Kervella}, {Girard}, {Kostogryz}, {Berdyugina}, {Perrin}, {Lacour},
  {Chiavassa}, \& {Ridgway}}]{houbois_2019}
{Haubois}, X., {Norris}, B., {Tuthill}, P.~G., {et~al.} 2019, \aap, 628, A101

\bibitem[{{Herzberg}(1948)}]{herzberg_1948}
{Herzberg}, G. 1948, \apj, 107, 94

\bibitem[{{Hjellming} \& {Newell}(1983)}]{hjellming_1983}
{Hjellming}, R.~M. \& {Newell}, R.~T. 1983, \apj, 275, 704

\bibitem[{{Josselin} {et~al.}(2000){Josselin}, {Blommaert}, {Groenewegen},
  {Omont}, \& {Li}}]{josselin_2000}
{Josselin}, E., {Blommaert}, J.~A.~D.~L., {Groenewegen}, M.~A.~T., {Omont}, A.,
  \& {Li}, F.~L. 2000, \aap, 357, 225

\bibitem[{{Kervella} {et~al.}(2011){Kervella}, {Perrin}, {Chiavassa},
  {Ridgway}, {Cami}, {Haubois}, \& {Verhoelst}}]{kervela_2011}
{Kervella}, P., {Perrin}, G., {Chiavassa}, A., {et~al.} 2011, \aap, 531, A117

\bibitem[{{Kudritzki} \& {Reimers}(1978)}]{kudritzki_1978}
{Kudritzki}, R.~P. \& {Reimers}, D. 1978, \aap, 70, 227

\bibitem[{{Lambert} {et~al.}(1984){Lambert}, {Brown}, {Hinkle}, \&
  {Johnson}}]{lambert_1984}
{Lambert}, D.~L., {Brown}, J.~A., {Hinkle}, K.~H., \& {Johnson}, H.~R. 1984,
  \apj, 284, 223

\bibitem[{{Lemaire} {et~al.}(2012){Lemaire}, {Vial}, {Curdt}, {Sch{\"u}hle}, \&
  {Woods}}]{lemaire_2012}
{Lemaire}, P., {Vial}, J.~C., {Curdt}, W., {Sch{\"u}hle}, U., \& {Woods}, T.~N.
  2012, \aap, 542, L25

\bibitem[{{Levesque}(2017)}]{levesque_2017}
{Levesque}, E.~M. 2017, {Astrophysics of Red Supergiants}

\bibitem[{{Levesque} {et~al.}(2005){Levesque}, {Massey}, {Olsen}, {Plez},
  {Josselin}, {Maeder}, \& {Meynet}}]{levesque_2005}
{Levesque}, E.~M., {Massey}, P., {Olsen}, K.~A.~G., {et~al.} 2005, \apj, 628,
  973

\bibitem[{{Lim} {et~al.}(1998){Lim}, {Carilli}, {White}, {Beasley}, \&
  {Marson}}]{lim_1998}
{Lim}, J., {Carilli}, C.~L., {White}, S.~M., {Beasley}, A.~J., \& {Marson},
  R.~G. 1998, \nat, 392, 575

\bibitem[{{Linsky}(2017)}]{linsky_2017}
{Linsky}, J.~L. 2017, \araa, 55, 159

\bibitem[{{Mart{\'{\i}}-Vidal} {et~al.}(2014){Mart{\'{\i}}-Vidal}, {Vlemmings},
  {Muller}, \& {Casey}}]{vidal_2014}
{Mart{\'{\i}}-Vidal}, I., {Vlemmings}, W.~H.~T., {Muller}, S., \& {Casey}, S.
  2014, \aap, 563, A136

\bibitem[{{Mathisen}(1984)}]{mathisen_1984}
{Mathisen}, R. 1984, Institute of Theoretical Astrophysics, University of Oslo,
  Publication Series 1

\bibitem[{{Matthews} {et~al.}(2015){Matthews}, {Reid}, \&
  {Menten}}]{matthews_2015}
{Matthews}, L.~D., {Reid}, M.~J., \& {Menten}, K.~M. 2015, \apj, 808, 36

\bibitem[{{Matthews} {et~al.}(2018){Matthews}, {Reid}, {Menten}, \&
  {Akiyama}}]{matthews_2018}
{Matthews}, L.~D., {Reid}, M.~J., {Menten}, K.~M., \& {Akiyama}, K. 2018, \aj,
  156, 15

\bibitem[{{McMurry}(1999)}]{mcmurry_1999}
{McMurry}, A.~D. 1999, \mnras, 302, 37

\bibitem[{{Mendoza} \& {Zeippen}(1987)}]{mendoza_1987}
{Mendoza}, C. \& {Zeippen}, C.~J. 1987, A\&A, 179, 346

\bibitem[{{Menten} {et~al.}(2012){Menten}, {Reid}, {Kami{\'n}ski}, \&
  {Claussen}}]{menten_2012}
{Menten}, K.~M., {Reid}, M.~J., {Kami{\'n}ski}, T., \& {Claussen}, M.~J. 2012,
  \aap, 543, A73

\bibitem[{{Meynet} {et~al.}(2015){Meynet}, {Chomienne}, {Ekstr{\"o}m},
  {Georgy}, {Granada}, {Groh}, {Maeder}, {Eggenberger}, {Levesque}, \&
  {Massey}}]{meynet_2015}
{Meynet}, G., {Chomienne}, V., {Ekstr{\"o}m}, S., {et~al.} 2015, \aap, 575, A60

\bibitem[{{Montarg{\`e}s} {et~al.}(2014){Montarg{\`e}s}, {Kervella}, {Perrin},
  {Ohnaka}, {Chiavassa}, {Ridgway}, \& {Lacour}}]{Montarges_2014}
{Montarg{\`e}s}, M., {Kervella}, P., {Perrin}, G., {et~al.} 2014, \aap, 572,
  A17

\bibitem[{{Nahar}(2000)}]{nahar_2000}
{Nahar}, S.~N. 2000, ApJS, 126, 537

\bibitem[{{Newell} \& {Hjellming}(1982)}]{newell_1982}
{Newell}, R.~T. \& {Hjellming}, R.~M. 1982, \apjl, 263, L85

\bibitem[{{O'Gorman} {et~al.}(2015){O'Gorman}, {Harper}, {Brown}, {Guinan},
  {Richards}, {Vlemmings}, \& {Wasatonic}}]{ogorman_2015}
{O'Gorman}, E., {Harper}, G.~M., {Brown}, A., {et~al.} 2015, \aap, 580, A101

\bibitem[{{O'Gorman} {et~al.}(2017){O'Gorman}, {Kervella}, {Harper},
  {Richards}, {Decin}, {Montarg{\`e}s}, \& {McDonald}}]{ogorman_2017}
{O'Gorman}, E., {Kervella}, P., {Harper}, G.~M., {et~al.} 2017, \aap, 602, L10

\bibitem[{{Ohnaka}(2004)}]{ohnaka_2004}
{Ohnaka}, K. 2004, \aap, 421, 1149

\bibitem[{{Ohnaka} {et~al.}(2013){Ohnaka}, {Hofmann}, {Schertl}, {Weigelt},
  {Baffa}, {Chelli}, {Petrov}, \& {Robbe-Dubois}}]{ohnaka_2013}
{Ohnaka}, K., {Hofmann}, K.-H., {Schertl}, D., {et~al.} 2013, \aap, 555, A24

\bibitem[{{Ohnaka} {et~al.}(2017){Ohnaka}, {Weigelt}, \&
  {Hofmann}}]{ohnaka_2017}
{Ohnaka}, K., {Weigelt}, G., \& {Hofmann}, K.~H. 2017, \nat, 548, 310

\bibitem[{{Ohnaka} {et~al.}(2011){Ohnaka}, {Weigelt}, {Millour}, {Hofmann},
  {Driebe}, {Schertl}, {Chelli}, {Massi}, {Petrov}, \& {Stee}}]{ohnaka_2011}
{Ohnaka}, K., {Weigelt}, G., {Millour}, F., {et~al.} 2011, \aap, 529, A163

\bibitem[{{Olnon}(1975)}]{olnon_1975}
{Olnon}, F.~M. 1975, \aap, 39, 217

\bibitem[{{Panagia} \& {Felli}(1975)}]{panagia_1975}
{Panagia}, N. \& {Felli}, M. 1975, \aap, 39, 1

\bibitem[{{Perrin} {et~al.}(2004){Perrin}, {Ridgway}, {Coud{\'e} du Foresto},
  {Mennesson}, {Traub}, \& {Lacasse}}]{perrin_2004}
{Perrin}, G., {Ridgway}, S.~T., {Coud{\'e} du Foresto}, V., {et~al.} 2004,
  \aap, 418, 675

\bibitem[{{Perrin} {et~al.}(2007){Perrin}, {Verhoelst}, {Ridgway}, {Cami},
  {Nguyen}, {Chesneau}, {Lopez}, {Leinert}, \& {Richichi}}]{perrin_2007}
{Perrin}, G., {Verhoelst}, T., {Ridgway}, S.~T., {et~al.} 2007, \aap, 474, 599

\bibitem[{{Planesas} {et~al.}(2016){Planesas}, {Alcolea}, \&
  {Bachiller}}]{planesas_2016}
{Planesas}, P., {Alcolea}, J., \& {Bachiller}, R. 2016, \aap, 586, A69

\bibitem[{{Pugh} \& {Gray}(2013)}]{pugh_2013}
{Pugh}, T. \& {Gray}, D.~F. 2013, \aj, 145, 38

\bibitem[{{Reid} \& {Menten}(1997)}]{reid_1997}
{Reid}, M.~J. \& {Menten}, K.~M. 1997, \apj, 476, 327

\bibitem[{{Reid} \& {Menten}(2007)}]{reid_2007}
{Reid}, M.~J. \& {Menten}, K.~M. 2007, \apj, 671, 2068

\bibitem[{{Reimers} {et~al.}(2008){Reimers}, {Hagen}, {Baade}, \&
  {Braun}}]{reimers_2008}
{Reimers}, D., {Hagen}, H.-J., {Baade}, R., \& {Braun}, K. 2008, \aap, 491, 229

\bibitem[{{Rodgers} \& {Glassgold}(1991)}]{rodgers_1991}
{Rodgers}, B. \& {Glassgold}, A.~E. 1991, \apj, 382, 606

\bibitem[{{Rybicki} \& {Lightman}(1986)}]{rybicki_1986}
{Rybicki}, G.~B. \& {Lightman}, A.~P. 1986, {Radiative Processes in
  Astrophysics}, 400

\bibitem[{{Ryde} {et~al.}(2006){Ryde}, {Harper}, {Richter}, {Greathouse}, \&
  {Lacy}}]{ryde_2006}
{Ryde}, N., {Harper}, G.~M., {Richter}, M.~J., {Greathouse}, T.~K., \& {Lacy},
  J.~H. 2006, \apj, 637, 1040

\bibitem[{{Sanner}(1976)}]{sanner_1976}
{Sanner}, F. 1976, \apjs, 32, 115

\bibitem[{{Schrijver} \& {Zwaan}(2000)}]{schrijver_2000}
{Schrijver}, C.~J. \& {Zwaan}, C. 2000, {Solar and Stellar Magnetic Activity}

\bibitem[{{Sim}(2001)}]{sim_2001}
{Sim}, S.~A. 2001, \mnras, 326, 821

\bibitem[{{Skinner} {et~al.}(1997){Skinner}, {Dougherty}, {Meixner}, {Bode},
  {Davis}, {Drake}, {Arens}, \& {Jernigan}}]{skinner_1997}
{Skinner}, C.~J., {Dougherty}, S.~M., {Meixner}, M., {et~al.} 1997, \mnras,
  288, 295

\bibitem[{{Smyth} {et~al.}(2019){Smyth}, {Ballance}, \&
  {Ramsbottom}}]{smyth_2019}
{Smyth}, R.~T., {Ballance}, C.~P., \& {Ramsbottom}, C.~A. 2019, \apj, 874, 144

\bibitem[{{Tsuji}(2000)}]{tsuji_2000}
{Tsuji}, T. 2000, \apj, 538, 801

\bibitem[{{Tsuji}(2006)}]{tsuji_2006}
{Tsuji}, T. 2006, \apj, 645, 1448

\bibitem[{{Uitenbroek} {et~al.}(1998){Uitenbroek}, {Dupree}, \&
  {Gilliland}}]{uitenbroek_1998}
{Uitenbroek}, H., {Dupree}, A.~K., \& {Gilliland}, R.~L. 1998, \aj, 116, 2501

\bibitem[{van Leeuwen(2007)}]{van_Leeuwen_2007}
van Leeuwen, F. 2007, A\&A, 474, 653–664

\bibitem[{{Visser} {et~al.}(2009){Visser}, {van Dishoeck}, \&
  {Black}}]{visser_2009}
{Visser}, R., {van Dishoeck}, E.~F., \& {Black}, J.~H. 2009, \aap, 503, 323

\bibitem[{{Vlemmings} {et~al.}(2019){Vlemmings}, {Khouri}, \&
  {Olofsson}}]{vlemmings_2019}
{Vlemmings}, W.~H.~T., {Khouri}, T., \& {Olofsson}, H. 2019, \aap, 626, A81

\bibitem[{{Wiedemann} {et~al.}(1994){Wiedemann}, {Ayres}, {Jennings}, \&
  {Saar}}]{wiedemann_1994}
{Wiedemann}, G., {Ayres}, T.~R., {Jennings}, D.~E., \& {Saar}, S.~H. 1994,
  \apj, 423, 806

\bibitem[{{Wood} {et~al.}(2005){Wood}, {Redfield}, {Linsky}, {M{\"u}ller}, \&
  {Zank}}]{wood_2005}
{Wood}, B.~E., {Redfield}, S., {Linsky}, J.~L., {M{\"u}ller}, H.-R., \& {Zank},
  G.~P. 2005, \apjs, 159, 118

\bibitem[{{Wright} \& {Barlow}(1975)}]{wright_1975}
{Wright}, A.~E. \& {Barlow}, M.~J. 1975, \mnras, 170, 41

\end{thebibliography}
\begin{appendix} 
\section{Searching for small-scale asymmetries}
\begin{figure*}[hbt!]
\centering 

\mbox{
\includegraphics[trim=0pt 0pt 0pt 0pt,clip, angle=90, scale=0.4]{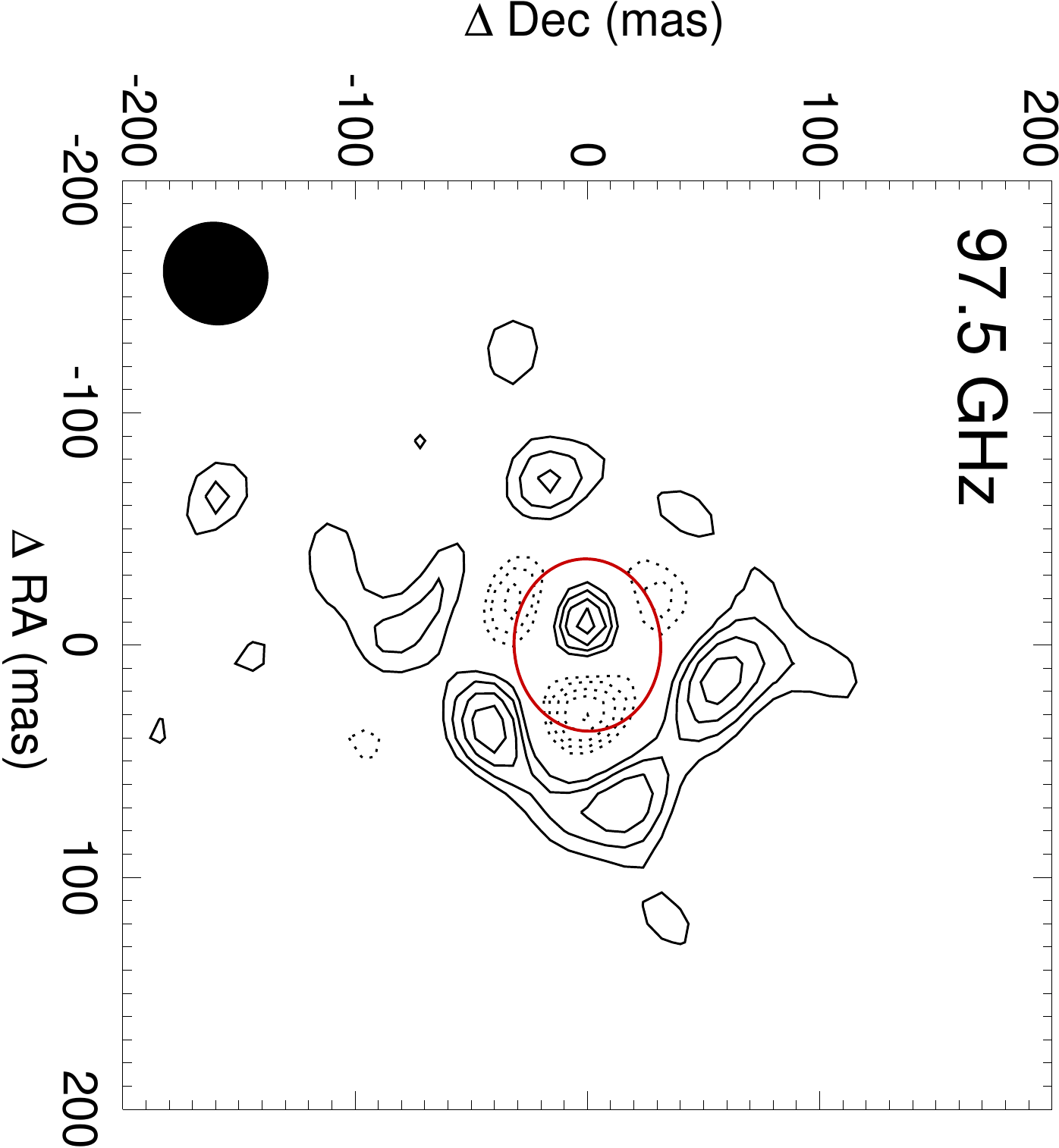}
\includegraphics[trim=0pt 0pt 0pt 0pt,clip, scale=0.4, angle=90]{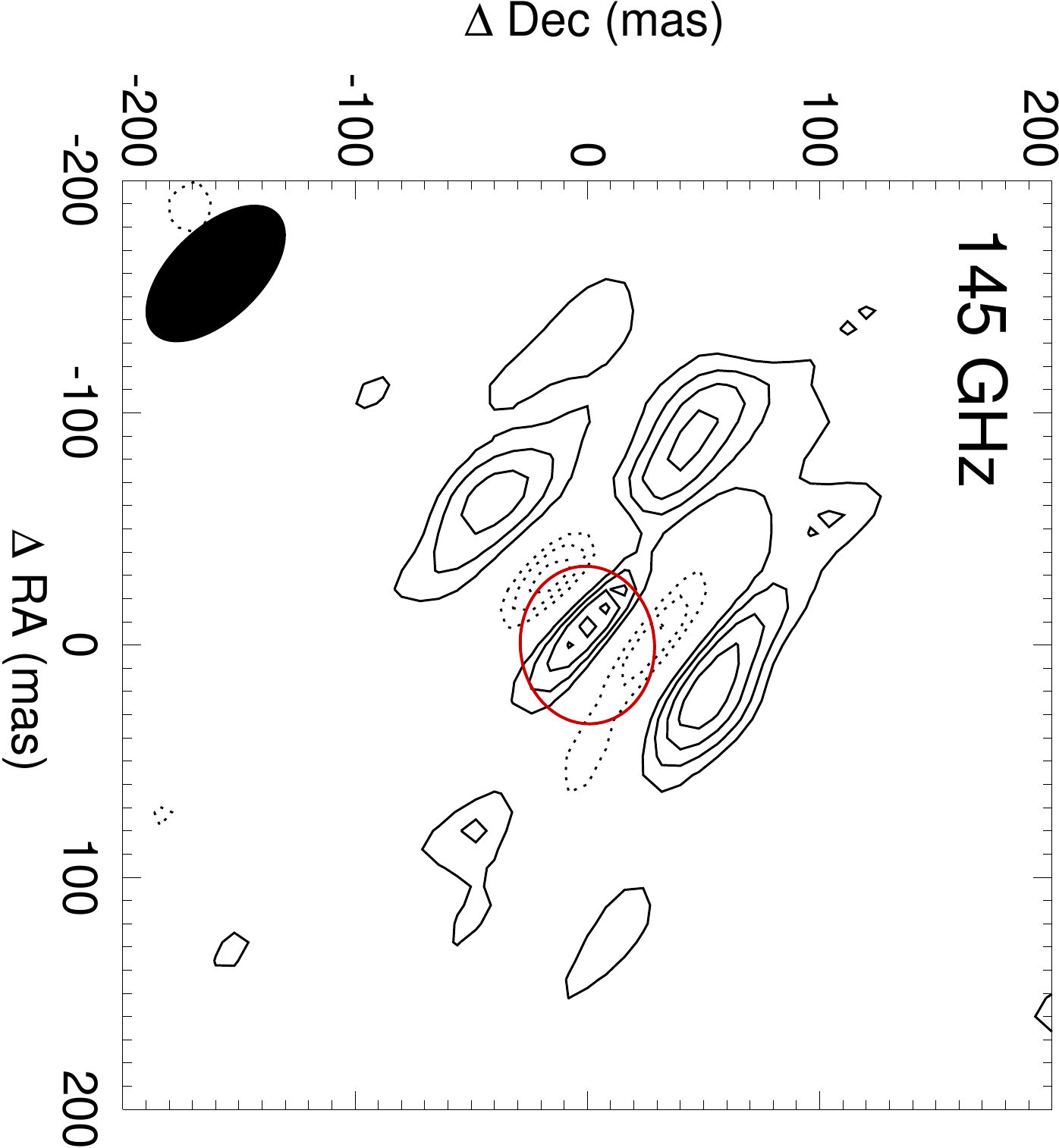}
}
\mbox{
\includegraphics[trim=0pt 0pt 0pt 0pt,clip,angle=90, scale=0.4]{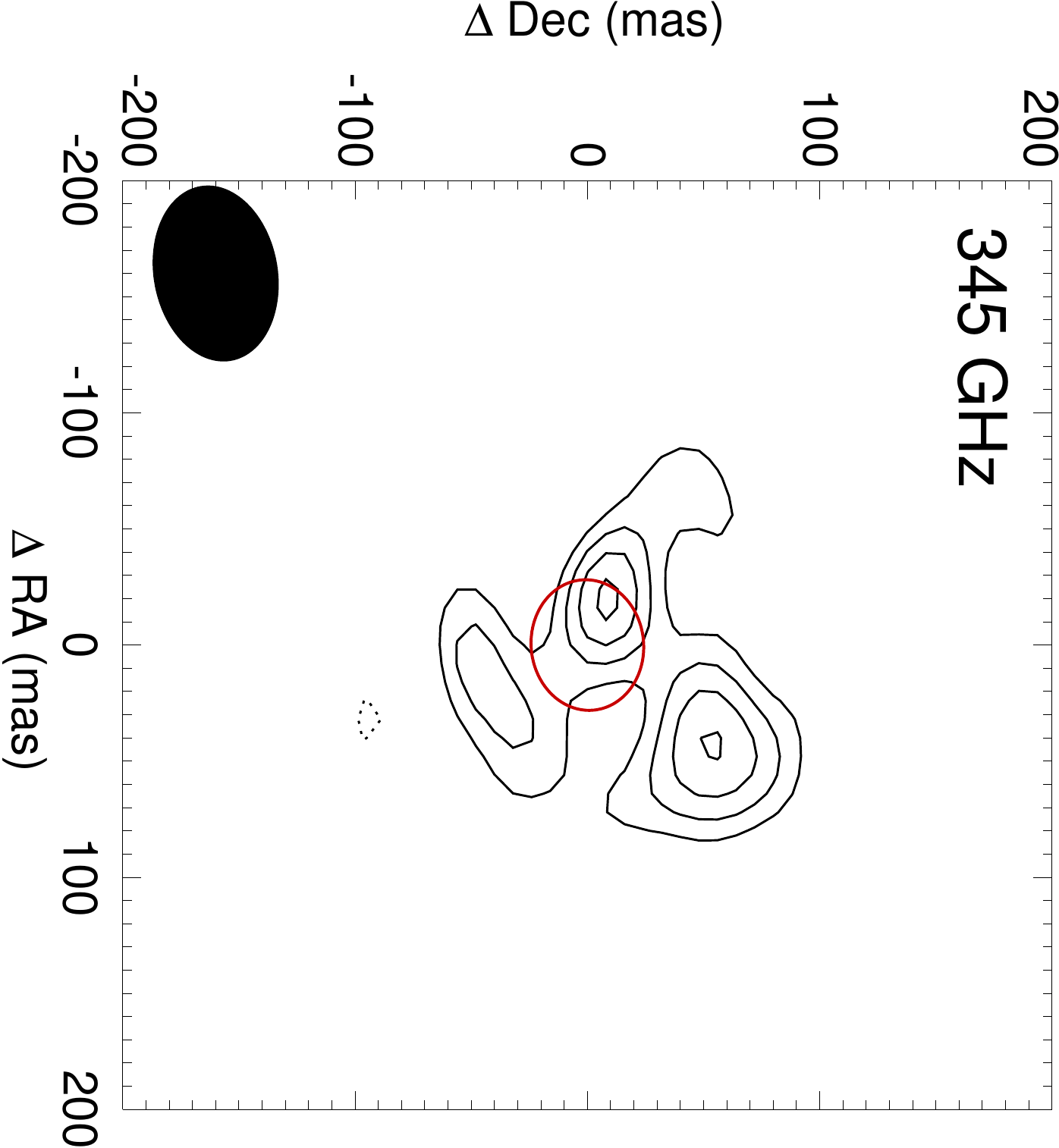}
\includegraphics[trim=0pt 0pt 0pt 0pt,clip, scale=0.4, angle=90]{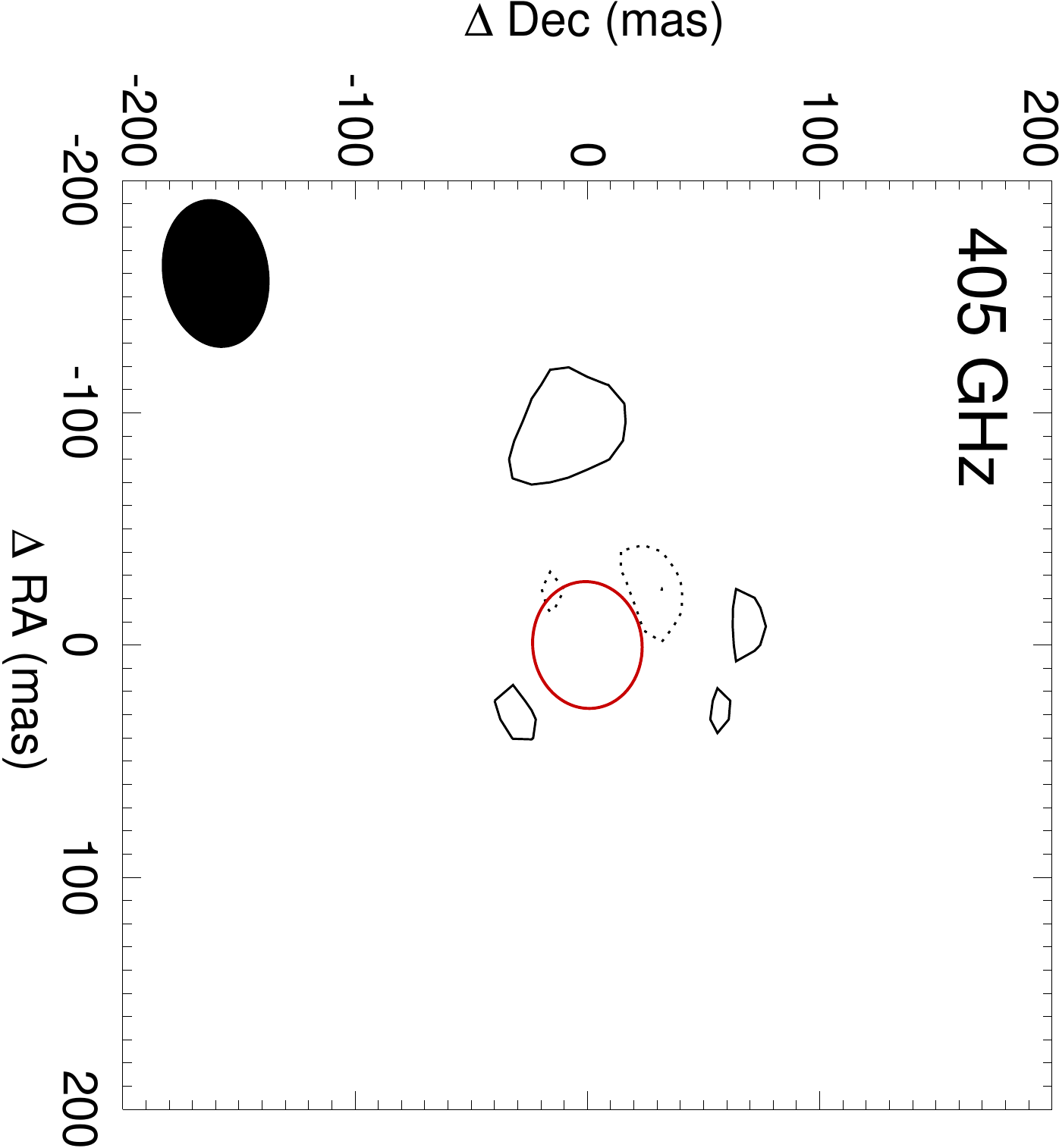}
}

\caption[]{ALMA Briggs-weighted (robust 0) residual images of Antares constructed from the visibilities from which the best-fit uniform-intensity elliptical disk models were first subtracted. The size and shape of the restoring beam at each frequency is shown in the lower left corner of each image. Contour levels are set to $(-10, -8, -6, -4, 4, 6, 8, 10)\times \sigma _{\textrm{rms}}$. The size and shape of the chromosphere is shown as a red ellipse in each image.}
\label{fig1a}
\end{figure*}

In Figure \ref{fig1a} we show the residual ALMA images where the best-fit uniform-intensity elliptical disks were first subtracted from the visibilities. Some weak ($\le$10\,$\sigma$) emission features are still present. However, we are unable to improve the visibility fitting by creating more complex models than those consisting of only uniform-intensity elliptical disks. The residual images indicate the possibility of small-scale deviations from uniformity in the emission throughout the chromosphere.  ALMA continuum data  with higher spatial resolution are required to detect such features, which are known to exist in the chromosphere of Betelgeuse \citep{ogorman_2017}.

\section{Calculating the optical depth of the MOLsphere and dust-forming gas}\label{app1}    
\cite{ohnaka_2013} estimated the CO column density and the radius of the MOLsphere of Antares to be $10^{20}$~cm$^{-2}$ and 1.3 R$_{\star}$ (i.e., 0.3 R$_{\star}$ above the photosphere), respectively. As the geometrical thickness of the MOLsphere, we adopted 0.3\,R$_{\star}$= $1.43 \times 10^{13}$~cm with R$_{\star}$ = 680 R$_{\odot}$. This gives a CO number density of $7 \times 10^{6}$~cm$^{-3}$. Assuming chemical equilibrium, we computed the gas pressure that reproduces this CO number density at 2000~K. We obtained a gas pressure of $\log P_{\rm gas} = -2.0$~dyn~cm$^{-2}$, 
an electron density of $\log n_e = 5.1$~cm$^{-3}$, a neutral hydrogen density of $\log n_{\rm H} = 10.5$~cm$^{-3}$, and a molecular hydrogen density of $\log n_{\rm H_2} = 8.0$~cm$^{-3}$. We only considered the opacity of H$^{-}$ because the neutral hydrogen density is much higher than that of molecular hydrogen. The formula given by \cite{dalgarno_1966} gives a cross section of $7.92 \times 10^{-21}$~cm$^4$~dyn$^{-1}$ at 0.08\,cm and 2000~K. Combined with the electron and neutral hydrogen number densities derived above, we obtained an optical depth of $1.3 \times 10^{-4}$ for the geometrical thickness of 0.3 R$_{\star}$. When we allow for photoionization of metals from the surrounding hot plasma, the electron density might be 2 orders of magnitude higher. Even in this case, the MOLsphere remains optically thin.

\section{Details of the radiative transfer model}\label{app2}
The \cite{harper_2001} semiempirical thermodynamic model of Betelgeuse was updated to include modern estimates of the distance \citep{harper_2017} and photospheric angular diameter. The original multiwavelength VLA angular diameters were supplemented with the ALMA 0.89\,mm measurement from \cite{ogorman_2017}. The main changes to the original model are therefore interior to the region of the peak radio temperature because the ALMA data now provide a direct temperature-radius measurement interior to the VLA values. This updated model will be published elsewhere.

For the simulations of the FUV continuum, the bound-free continua of Si~I, Fe~I, C~I, and Mg~I were solved simultaneously using three-term model ions, with the first two terms of the neutral state and the ground term of the singly ionized atom. Collisional and radiative transitions were included between all terms. Background opacities from Ca~I, Al~I, and S~I were included using a photoionization-radiative recombination approximation appropriate for the observed low gas temperatures. The adopted photoionization cross sections are for Si I \citep{nahar_2000}, Fe I \citep{smyth_2019}, and Mg I \citep{mendoza_1987}, Ca I and Al I (from the compilation of \citealt{mathisen_1984}), and H I Rayleigh-scattering cross sections \citep{gavrila_1967}. The photoionization cross sections for C I and S I were taken from fits made by P. D. Bennett (private communication) based on Topbase values \citep{cunto_1993}. The hydrogen
ionization was treated using the escape probability approximation
given in \cite{hartmann_1984}. For the enhanced
photoionization simulation, the Ly\,$\alpha$ to Ly\,$\beta$ flux ratio
was assumed to be 250, the value found from hydrogen partial redistribution computations for the inactive red giant $\alpha$~Tau (K5~III) by \cite{sim_2001}. This value is slightly higher than that found in quiet areas of the Sun \citep{lemaire_2012}. The abundances for Betelgeuse were taken from \cite{lambert_1984}, \cite{rodgers_1991}, and \cite{carr_2000}, and any other elements were taken as solar values  \citep{asplund_2009}.

The radiative transfer problem for the bound-free continua of Si, Fe, C, and Mg was solved simultaneously for 1000 frequency points
for 960 < $\lambda$(Å) < 2060 for $1 < R_\star < 15$ using 251 shells and 15 core rays using standard angle quadrature (e.g., \citealt{harper_1994}). The inner boundary condition was a MARCS photospheric model \citep{gustafsson_2008}. The ion densities of
Si, Fe, C, and Mg, were iterated until convergence. A description of the full complexities of detailed FUV solar simulations can be found in \cite{avrett_2008} and \cite{fontenla_2011} and references therein.

\end{appendix}
\end{document}